\documentclass{nature}

\usepackage{pdfpages}
\usepackage{setspace}
\usepackage{amsfonts}
\usepackage{amssymb}
\usepackage{amsmath}

\usepackage{float}
\usepackage{placeins}
\usepackage{graphicx}

\usepackage[export]{adjustbox}

\usepackage{graphicx,graphics,epsfig,subfigure,times,bm,bbm,amssymb,amsmath,amsthm,mathrsfs,MnSymbol} \usepackage{gensymb} \usepackage{amsfonts} \usepackage{float} \usepackage[matrix,frame,arrow]{xypic} \usepackage[pdfstartview=FitH]{hyperref} 
\usepackage{times}
\usepackage{float}
\usepackage{graphics}
\usepackage[T1]{fontenc}
\usepackage{braket} 
\usepackage[pdfstartview=FitH]{hyperref} 

\hypersetup{ colorlinks=true,       
	linkcolor=red,          
	citecolor=blue,        
	filecolor=magenta,      
	urlcolor=blue,           
	runcolor=cyan }

\title{Characterizing Many-body Dynamics with Projected Ensembles on a Superconducting Quantum  Processor}

\author{Zhiguang~Yan$^{1,*}$, Zi-Yong~Ge$^{1,*}$, Rui~Li$^{1}$, Yu-Ran~Zhang$^{2}$, Franco~Nori$^{1,3,\dagger}$ and Yasunobu~Nakamura$^{1,4,\ddagger}$}

\begin{document}
	
	\maketitle
	
	\begin{affiliations}
 \item RIKEN Center for Quantum Computing (RQC), Wako, Saitama 351-0198, Japan
\item School of Physics and Optoelectronics, South China University of Technology, Guangzhou 510640, China
\item Department of Physics, University of Michigan, Ann Arbor, Michigan 48109-1040, USA
\item  Department of Applied Physics, Graduate School of Engineering, The University of Tokyo, Bunkyo-ku, Tokyo 113-8656, Japan
\item[] $^{*}$These authors contributed equally to this work.
 \item[] $^{\dagger}$ fnori@riken.jp
  \item[] $^{\ddagger}$ yasunobu@ap.t.u-tokyo.ac.jp
	\end{affiliations}
	
	\begin{abstract}   
    Quantum simulators~\cite{buluta2009quantum,trabesinger2012quantum,RevModPhys.86.153} offer a new opportunity for the experimental exploration of non-equilibrium quantum many-body dynamics~\cite{RevModPhys.83.863}, which have traditionally been characterized through expectation values or entanglement measures, based on density matrices of the system.
    Recently, a more general framework for studying quantum many-body systems based on projected ensembles~\cite{choi2023preparing,PhysRevLett.128.060601,PRXQuantum.4.010311,PhysRevX.14.041051,ge2024identifying}
    has been introduced, revealing novel quantum phenomena, 
    such as deep thermalization in chaotic systems~\cite{choi2023preparing,PhysRevLett.128.060601,PRXQuantum.4.010311}. 
    Here, we experimentally investigate 
    a chaotic quantum many-body system using projected ensembles on a superconducting processor~\cite{koch2007charge,annurev-conmatphys-031119-050605,siddiqi2021engineering} with 16 qubits on a square lattice. Our results provide direct evidence of deep thermalization by observing a Haar-distributed projected ensemble for the steady states within a charge-conserved sector. 
    Moreover, by introducing an ensemble-averaged entropy as a metric, we establish a benchmark for many-body information leakage from the system to its environment.
    Our work paves the way for studying quantum many-body dynamics using projected ensembles and shows a potential implication
    for advancing quantum simulation techniques.

\end{abstract}

Understanding the non-equilibrium dynamics of closed quantum many-body systems~\cite{RevModPhys.83.863} remains a central challenge in modern quantum physics. 
Recent advances in quantum simulations~\cite{buluta2009quantum,trabesinger2012quantum,RevModPhys.86.153} have provided powerful experimental platforms to explore this problem in depth. 
It has been shown that isolated chaotic systems, initialized in typical pure states, can evolve toward thermal equilibrium, where the reduced density matrix of a local subsystem resembles a Gibbs ensemble~\cite{PhysRevA.43.2046,PhysRevE.50.888,Srednicki_1999,Rigol2008Nature,DAlessio2016}. 
This process, known as quantum thermalization, is well described by the Eigenstate Thermalization Hypothesis (ETH)~\cite{Srednicki_1999}.
However, reduced density matrices capture only limited information, notably lacking access to higher-order moments of the quantum ensemble. 
To address this limitation, the concept of \textit{projected ensembles} has recently been introduced~\cite{choi2023preparing,PhysRevLett.128.060601,PRXQuantum.4.010311,PhysRevX.14.041051}, providing a more complete characterization of quantum thermalization. 
A projected ensemble is constructed by performing projective measurements on the complementary subsystem in a fixed local basis, yielding a collection of post-measurement wavefunctions on the subsystem of interest. 
Thus, projected ensembles retain full information about the many-body wavefunction and can uncover universal properties beyond what is accessible only through reduced density matrices.
In particular, in certain chaotic systems, projected ensembles can exhibit the emergence of quantum state designs~\cite{renes2004symmetric,ambainis2007quantum} after long-time evolution, a phenomenon known as \textit{deep thermalization}~\cite{choi2023preparing,PhysRevLett.128.060601,PRXQuantum.4.010311}, 
which is a stronger form of quantum thermalization. 
Despite its theoretical importance, direct experimental observation of deep thermalization has remained elusive.

In quantum simulations, another fundamental concern is evaluating how well a quantum system remains isolated from its environment, 
which requires quantifying the information leakage. 
Conventionally, single-qubit energy relaxation time $T_1$ and dephasing time $T^*_2$ are used to characterize the coupling strength to the environment. 
However, $T_1$ and $T^*_2$ are insufficient to accurately capture information leakage in many-body regimes. 
For example, it has been shown that dephasing can be suppressed when all qubits in a superconducting circuit are tuned to resonance~\cite{PhysRevLett.116.010501}.
A general approach to quantify quantum information leakage is through entropy measures, which are extremely challenging to access in quantum many-body systems~\cite{PhysRevLett.120.050507,brydges2019probing,google2023measurement,WOS:001207800900011}.
This highlights the need for scalable methods to benchmark \textit{many-body information leakage} in quantum simulators. 
While measuring the full projected ensemble still remains experimentally difficult, to study a subset of high-probability states within the projected ensemble is accessible for experiments. 
This reduced ensemble still encodes substantial global information about the underlying quantum many-body wavefunction.
These considerations naturally lead to an intriguing question: Can projected ensembles offer a practical and scalable approach to benchmarking many-body information leakage in quantum simulators?


In this work, using a 16-qubit superconducting quantum processor, we experimentally investigate the dynamics of a quantum many-body system from the perspective of projected ensembles.
The effective Hamiltonian of the system is described by a 2D spin-$\frac{1}{2}$ XY model, exhibiting spin $U(1)$ symmetry. 
Focusing on a subsystem of two nearest-neighbor qubits, 
we explore deep thermalization starting from a half-filling product state.
After a long-time evolution, the projected ensemble within the spin-conserved sector exhibits a Haar-random distribution,
providing direct experimental evidence of deep thermalization.
Furthermore, by using ensemble-averaged entropies of projected ensembles,
we benchmark many-body information leakage, which can hardly be captured by single-qubit decoherence measurements.
Our results highlight the power of projected ensembles in characterizing quantum many-body dynamics,
offering insights beyond conventional observables based on density matrices.

\section*{Projected ensembles and experimental setup}

We begin by briefly reviewing the concept of projected ensembles.
We partition a quantum many-body system into two parts: a local subsystem  A and its complement B. 
For a quantum state $\ket{\Psi}$, the wave function can be expressed as:
$\ket{\Psi} = \sum _{z_\mathrm{B}}\sqrt{p(z_\mathrm{B})}\ket{\Psi_{\mathrm{A}}(z_\mathrm{B})}\otimes \ket{z_\mathrm{B}}$,
where $ \ket{z_{\mathrm{B}}} $ is a measurement basis for  B  (represented as a bit-string in our experiment),  $p(z_\mathrm{B})$  is the probability of measuring the bit-string  $z_\mathrm{B}$, and $ \ket{\Psi_{\mathrm{A}}(z_\mathrm{B})} $ is the resulting state of  A  when  B  collapses to $ \ket{z_\mathrm{B}}$ (Fig.~\ref{fig_1}a). The projected ensemble of subsystem A with respect to $ \ket{\Psi}$ is then defined as:
\begin{align} 
	\mathcal{E}_{\Psi, \mathrm{A}} := \{p(z_\mathrm{B}),\ket{\Psi_{\mathrm{A}}(z_\mathrm{B})}\}.
\end{align}	
Here,  $\mathcal{E}_{\Psi, \mathrm{A}}$ captures all  information about the quantum many-body state $\ket{\Psi}$, superior to the reduced density matrix of  A.
In an infinite-temperature system without any conserved charges, deep thermalization is characterized by the emergence of a quantum state design~\cite{choi2023preparing,PhysRevLett.128.060601,PRXQuantum.4.010311,PhysRevX.14.041051}, i.e., $\mathcal{E}_{\Psi, \mathrm{A}}$ tends towards a Haar ensemble after a long-time evolution (Fig.~\ref{fig_1}b). 

Our experiment is performed on a scalable and 3D-integrated 
superconducting quantum processor, which consists of 16 concentric and frequency-tunable transmon qubits~\cite{koch2007charge,annurev-conmatphys-031119-050605,siddiqi2021engineering} arranged in a  $4 \times 4$  square lattice (Fig.~\ref{fig_1}c, Methods). 
The qubits and readout circuits are patterned on the topside of the chip, while the contact pads for the control and readout ports are located on the backside. Those ports are connected to coaxial cables via spring contacts, enabling vertical wiring in a scalable manner. In addition, numerous superconductor-metalized through-silicon vias (TSVs) are placed across the chip. These TSVs have several crucial functions (see the Supplementary Information for more details): connecting the ground planes on the topside and backside of the chip, localizing control and readout signals for crosstalk suppression, suppressing substrate modes, and working as transmission lines for the readout. 
The median energy-relaxation time $T_1$ of these qubits is $43$~\textmu s, 
and the average readout fidelities are about $99.6\%$ and $97.5\%$ for states $\ket{0}$ and $\ket{1}$, respectively.

During the time evolution, all qubits are tuned in resonance with each other.
Thus, the effective Hamiltonian of this system can be described by a 2D spin-$\frac{1}{2}$ XY model~\cite{roushan2017spectroscopic,yan2019strongly,PhysRevLett.123.050502}, which is expressed as ($\hbar = 1$)
\begin{align} \label{H}
	\hat H = \sum _{<i,j>}J_{ij}(\hat\sigma_i^+\hat\sigma_j^-+\mathrm{h.c.}),
\end{align}	
where $\hat\sigma^\alpha $ (for $\alpha\in\{x,y,z\}$) are the Pauli matrices, and the nearest-neighbor coupling $J_{ij}$ is about $2\pi\times4$~MHz (Fig.~\ref{fig_1}d).
In the Supplementary Information,  the detailed parameters of the device are presented.
The Hamiltonian $\hat H$ is a typical chaotic system with spin $U(1)$ symmetry~\cite{WOS:001207800900011,andersen2025thermalization}, 
i.e., the total spin $\sum_j \hat\sigma_j^z$ is conserved.

\section*{Observation of deep thermalization}  
Here, we utilize projected ensembles to study deep thermalization.
We select a half-filling state $\ket{\Psi_0} = \ket{0101...01}$ as the initial state. In this configuration, $\braket{\Psi_0|\hat H|\Psi_0}=0$, corresponding to an infinite-temperature system. 
In addition, we select two bulk qubits,  Q$_5$  and  Q$_6$, as the subsystem  A,
while other qubits serve as the subsystem B (Fig.~\ref{fig_1}d).
In our experiment, we first prepare the initial state using single-qubit rotation gates on the target qubits and then bring all qubits into resonance. After a time  $t$, the system evolves to the state $\ket{\Psi(t)}=e^{-i\hat H t/\hbar }\ket{\Psi_0}$. We finally measure the relevant observables using joint single-shot readouts (Fig.~\ref{fig_1}e).

First,  the ergodicity of the system is examined by monitoring a local observable,  the density of qubit excitations $n_j := \bra{\Psi(t)}\hat\sigma_j^+\hat\sigma_j^-\ket{\Psi(t)}$.
As the system evolves, the distribution of  $n_j$  becomes homogeneous (Fig.~\ref{fig_2}a), which is a signature of conventional quantum thermalization.
Furthermore, we analyze the statistics of the bit string probability, 
$p(z)=|\braket{z|\Psi(t)}|^2$, where  $z$  is the outcome of the measurement bit-string of the entire system. 
In Fig.~\ref{fig_2}b, we show the distributions of  $p(z)$  at different evolution times. The results indicate that, for a large  $t$,  $p(z)$  follows the Porter-Thomas distribution 
$P(p)=\mathcal{D} e^{-\mathcal{D}p}$~\cite{andersen2025thermalization,PhysRev.104.483,WOS:000492991700045,WOS:000980800100012}, 
where  $\mathcal{D}$ is the dimension of the Hilbert space. 
We also consider the statistics of the  measured conditional probability of subsystem A~\cite{choi2023preparing}, written as $p(z_\mathrm{A}|z_\mathrm{B})$.
The experimental results show that $p(z_\mathrm{A}=10|z_\mathrm{B})$ exhibits a uniform distribution for a large $t$ (Fig.~\ref{fig_2}c).
These results indicate that the steady state is well described by a random state, indicating the ergodicity of the system.

To study deep thermalization, we also need to consider projected ensembles. 
The projected ensemble of $\ket{\Psi(t)}$  is measured using joint measurements with single-shot readout (Fig.~\ref{fig_1}e). 
For subsystem  B, the measurement is performed on the  $z$-basis with outcomes  $z_\mathrm{B}$. 
For the two-qubit subsystem A (Q$_5$  and  Q$_6$), we perform quantum state tomography by applying appropriate rotation gates to adjust the measurement basis before readout (Methods).
Due to the conservation of total spin, the projected state $\ket{\Psi_{\mathrm{A}}(z_\mathrm{B})}$  is the eigenstate of  $\hat{\sigma}_5^z + \hat{\sigma}_6^z$, which prevents the projected ensemble  $\mathcal{E}_{\Psi(t), \mathrm{A}}$  from forming a Haar distribution. However, we can focus on the half-filling sector of the projected ensembles:
$\mathcal{E}^{\text{hf}}_{\Psi(t), \mathrm{A}} := \{p(z_\mathrm{B}|_{n(z_\mathrm{B})=7}),\ket{\Psi_{\mathrm{A}}(t,z_\mathrm{B})}\}$,
where  $n(z_\mathrm{B})$  denotes the total excitations in $ z_\mathrm{B}$. In this scenario, the output state of subsystem  A  is given by $ \ket{\Psi_{\mathrm{A}}(t,z_\mathrm{B})} = \alpha(t,z_\mathrm{B})\ket{01} + \beta(t,z_\mathrm{B})\ket{10}$. 
Thus, if  $\mathcal{E}^{\text{hf}}_{\Psi, \mathrm{A}}$  approaches a Haar ensemble, it indicates that deep thermalization has occurred in this  $U(1)$-symmetric system~\cite{PhysRevX.14.041051}.

Due to the system's coupling with the environment, the output states $\ket{\Psi_{\mathrm{A}}(t,z_\mathrm{B})}$ are often mixed and may have nonzero components in the $\ket{00}$  and $\ket{11}$  bases. Therefore, we use the observed density matrices  $\hat{\rho}_{\mathrm{A}}(t, z_\mathrm{B})$  instead of  $\ket{\Psi_{\mathrm{A}}(t, z_\mathrm{B})}$. To further mitigate the impact of noise, we perform post-selection by projecting  $\hat{\rho}_{\mathrm{A}}(t, z_\mathrm{B})$  onto the subspace spanned by $\ket{01}$ and $\ket{10}$ as
$\tilde{ \rho}_{\mathrm{A}}(t,z_\mathrm{B}) = {\hat\Pi \hat \rho_{\mathrm{A}}(t,z_\mathrm{B}) \hat\Pi}/{\text{Tr}(\hat\Pi \hat \rho_{\mathrm{A}}(t,z_\mathrm{B}) \hat\Pi)}$,
where $\hat \Pi =\ket{01}\bra{01} +\ket{10}\bra{10}$.
The distributions of  $\tilde{ \rho}_{\mathrm{A}}(t,z_\mathrm{B})$  on the Bloch sphere are shown in Fig.~\ref{fig_3}a. At early times,  $\tilde{ \rho}_{\mathrm{A}}(t,z_\mathrm{B})$  is primarily concentrated in a localized region. As the system evolves,  $\tilde{ \rho}_{\mathrm{A}}(t,z_\mathrm{B})$ gradually delocalizes, and then becomes nearly uniform over the Bloch sphere. 
This provides strong evidence that  $\mathcal{E}^{\text{hf}}_{\Psi(t), \mathrm{A}}$ approximates a Haar ensemble after a sufficiently long time evolution, which cannot be identified by the density matrix.

We also introduce the  $k$th-moment density matrix of  $\mathcal{E}^{\text{hf}}_{\Psi(t), \mathrm{A}}$ as
$\tilde{ \rho}^{(k)}_{\mathrm{A}}(t) = \sum_{z_\mathrm{B}} p(z_\mathrm{B})[\tilde{ \rho}_{\mathrm{A}}(z_\mathrm{B})]^{\otimes k}$,
where $\tilde{ \rho}^{(1)}_{\mathrm{A}}$ is the density matrix of subsystem A.
Generally, two ensembles are considered equivalent if their  $k$th-moment density matrices are identical for any  $k$. In Fig.~\ref{fig_3}b, we present the density matrices $\tilde{ \rho}^{(2)}_{\mathrm{A}}$ and $\tilde{ \rho}^{(3)}_{\mathrm{A}}$  at  $t = 306$~ns, which approximate the corresponding $k$th-moment density matrices of the Haar ensemble.
To quantitatively evaluate the distance between $\mathcal{E}^\text{hf}_{\Psi(t), 
\mathrm{A}}$ and the Haar ensemble, we define the trace distance between the 
$k$th-moment density matrix of  $\mathcal{E}^{\text{hf}}_{\Psi(t), \mathrm{A}}$ and that of the Haar ensemble
$ \Delta^{(k)} := \frac{1}{2} ||\tilde{\rho}^{(k)}_{\mathrm{A}} -  \hat{\rho}^{(k)}_{\text{Haar}}||$,
where $||\cdot||$ is the trace norm.
As shown in Fig.~\ref{fig_3}c, $\Delta^{(k)}$ approaches a small value after a long time $t\gtrsim100$~ns,
providing direct experimental evidence of deep thermalization.
Note that, due to finite-size effects and intrinsic noise, $\Delta^{(k)}$ does not vanish completely.

We now turn to the impact of noise on deep thermalization.
The inherent coupling between the system and its environment can degrade the purity of quantum states,
thereby altering the distribution of projected ensembles.
To quantify this effect, we introduce the entropy of the $k$th-moment density matrix, defined as     
$S_\mathrm{A}^{(k)} := \text{Tr}(\tilde{\rho}^{(k)}_{\mathrm{A}}\ln \tilde{\rho}^{(k)}_{\mathrm{A}})$.
If the entire system is in a pure state, we have  $S_\mathrm{A}^{(k)} \leq \ln(k+1)$, with equality holding for a Haar ensemble. 
As shown in Fig.~\ref{fig_3}d, when $k > 1$, the value of $S_\mathrm{A}^{(k)}$ goes beyond $\ln(k+1)$ 
after a certain evolution time~($t\gtrsim100$~ns), 
indicating that the state of the entire system becomes mixed.
Thus, coupling with the environment prevents the projected ensemble from fully approaching a Haar ensemble. 
Nevertheless, Fig.~\ref{fig_3}d also demonstrates that the higher-order moments of the projected ensemble 
are highly sensitive to purity, 
suggesting their potential utility in detecting information leakage for the system to its environment.

\section*{Benchmarking many-body information leakage}   
Understanding environment-induced quantum coherence decay and relaxation is essential for the development of quantum computation and quantum simulation~\cite{lidar2019lecture}. 
However, conventional coherence measures, such as entropy, are difficult to apply in large-scale quantum simulators. 
Therefore, it is crucial to develop scalable approaches for quantifying many-body information leakage. 
We have shown that projected ensembles can capture purity of quantum many-body wavefunctions, 
making them a promising tool for assessing many-body information leakage.

We now utilize projected ensembles to benchmark many-body information leakage.
For a projected ensemble  $\mathcal{E}_{\Psi, \mathrm{A}}$, 
the information leakage makes a state $\hat{\rho}_\mathrm{A}(z_\mathrm{B})$ become mixed.
Intuitively, as entropy of the entire system increases, the purity of states in  $\mathcal{E}_{\Psi, \mathrm{A}}$ decreases. 
To quantify this, we introduce an ensemble-averaged entropy, defined as
\begin{align} \label{Ea}
	\bar E_A =  -\sum_{z_\mathrm{B}} p(z_\mathrm{B})\ln\text{Tr}\hat \rho_{\mathrm{A}}^2(z_\mathrm{B}).
\end{align}
A larger  $\bar{E}_\mathrm{A}$  indicates a lower purity of the system and a larger information leakage to the environment. 
In the Supplementary Information, we demonstrate that  $\bar{E}_\mathrm{A}$ is generally proportional to the entropy of the entire system for small information leakage. 
In this case,  $\bar{E}_\mathrm{A}$  can be approximated by a linear function (Supplementary Information):
\begin{align} \label{Ea}
	\bar E_\mathrm{A} \approx E_0 \frac{t}{\tau_\mathrm{MB}},
\end{align}
where  $E_0$ represents the steady-state value of  $\bar{E}_\mathrm{A}$, and  $\tau_{\text{MB}}$ serves as a decoherence time of the quantum many-body system, characterizing the speed of information leakage.

We first consider a $3 \times 3$ qubit system, and the central qubit (Q$_6$) is designated as subsystem A (inset of Fig.~\ref{fig_4}a). The initial state is prepared as $\ket{\psi_1} = \bigotimes_{j=\text{even}} \ket{X_+} \bigotimes_{j=\text{odd}} \ket{Y_+}$, where $\ket{X_+}$ and $\ket{Y_+}$ are the eigenstates of $\hat{\sigma}^x$ and $\hat{\sigma}^y$, respectively, with eigenvalue $+1$. 
In this system, $T_1$ is much longer than $T_2^*$ and the evolution time, 
indicating that the effect of $T_1$ can be neglected.
The source for dephasing can generally be categorized into two types: white noise and $1/f$ noise~\cite{RevModPhys.86.361}.
To study their effects, we perform numerical simulations of $\bar{E}_\mathrm{A}$ under white and $1/f$ noise, respectively, using the measured $T_2^*$ values at the operation point (Methods). 
The experimental and numerical results for $\bar{E}_\mathrm{A}$ are compared in Fig.~\ref{fig_4}a. 
By fitting the experimental results with Eq.~(\ref{Ea}) ($E_0=\ln 2$), we extract a lifetime of 0.94~\textmu s 
(The observed offset may result from
this initial state's sensitivity to dephasing noise during the Z-pulse tuning from the idle to the operation point.),
while the numerical predictions yield lifetimes of 0.40~\textmu s and 4.5~\textmu s for white noise and $1/f$ noise, respectively. 

Our numerical results indicate that $1/f$ noise leads to less information leakage compared to white noise, suggesting that quantum many-body systems are less sensitive to $1/f$ noise. This can be understood as follows: $1/f$ noise is the dominant source of dephasing in superconducting qubits~\cite{PhysRevLett.97.167001,4757203,siddiqi2021engineering}. When all qubits are tuned to be resonant, their rapid coherent dynamics effectively average out the low-frequency fluctuations of $1/f$ noise, thereby suppressing decoherence—a mechanism reminiscent of motional narrowing in nuclear magnetic resonance systems.
However, in multi-qubit superconducting circuits, additional non-negligible noise sources may exist, such as leakage to higher excited states or coupling to spurious two-level systems. As a result, the experimentally observed lifetime falls between the lifetimes predicted for white noise and $1/f$ noise, reflecting the combined influence of various noise channels.

We further extend our analysis to a 16-qubit system with spin-conserving, half-filling initial states [Fig.\ref{fig_4}(b)].
The experimentally observed $\bar{E}_\mathrm{A}$ again shows a linear increase at early times and lies between the numerical predictions for white and $1/f$ noise. 
These findings are consistent with those in the spin-non-conserving case [Fig.\ref{fig_4}(a)].


Due to the varying sensitivities of quantum many-body systems to different types of noise, single-qubit dephasing models are insufficient to fully capture the nature of information leakage in quantum many-body systems. 
Therefore, employing $\bar{E}_\mathrm{A}$ as a benchmark for many-body information leakage is crucial for evaluating the performance of noisy quantum simulators. 
Although measuring $\bar{E}_\mathrm{A}$ remains exponentially challenging for large-scale systems, this limitation can be mitigated by partitioning the system into multiple medium-sized blocks and detecting the information leakage within each block individually. 
This strategy can still offer insights beyond what single-qubit analyses can provide, 
enabling a more comprehensive characterization of noise-induced effects in large quantum systems.
Another promising application of projected ensembles in the context of open quantum systems~\cite{lidar2019lecture} 
lies in characterizing information flow and non-Markovianity~\cite{RevModPhys.88.021002}.
Conventional measures of these phenomena, such as entanglement or trace distance~\cite{PhysRevLett.105.050403,PhysRevA.82.042103,PhysRevLett.124.210502}, are difficult to scale to many-body regimes. 
In contrast, projected ensembles provide a potentially scalable alternative, 
making it a valuable tool for probing the non-Markovian dynamics and information backflow in open quantum many-body systems.

\section*{Conclusion}  
In summary, we have experimentally explored deep thermalization and benchmarked many-body information leakage using projected ensembles on a 2D superconducting circuit. Our results provide a direct experimental observation for the existence of deep thermalization. Moreover, we demonstrate that ensemble-averaged entropies can effectively quantify the extent of information leakage from a quantum many-body system to its environment,
and  reveal that single-qubit decoherence and dephasing times fail to accurately characterize the impacts of noise on quantum many-body systems.
Our findings demonstrate that projected ensembles offer a powerful approach for probing quantum many-body dynamics, providing complementary insights that go beyond those accessible through conventional density-matrix-based observables.
Thus, our approach holds significant potential for advancing quantum simulation.
Moreover, since projected ensembles can capture global information of quantum many-body wavefunctions, 
they may find broader applications in deepening our understanding of other quantum many-body physics, 
including open quantum many-body systems~\cite{lidar2019lecture}, 
topological phases~\cite{PhysRevLett.96.110404,PhysRevB.82.155138}, 
and entanglement phases~\cite{ge2024identifying}.

\begin{figure}[t] \includegraphics[width=1\textwidth]{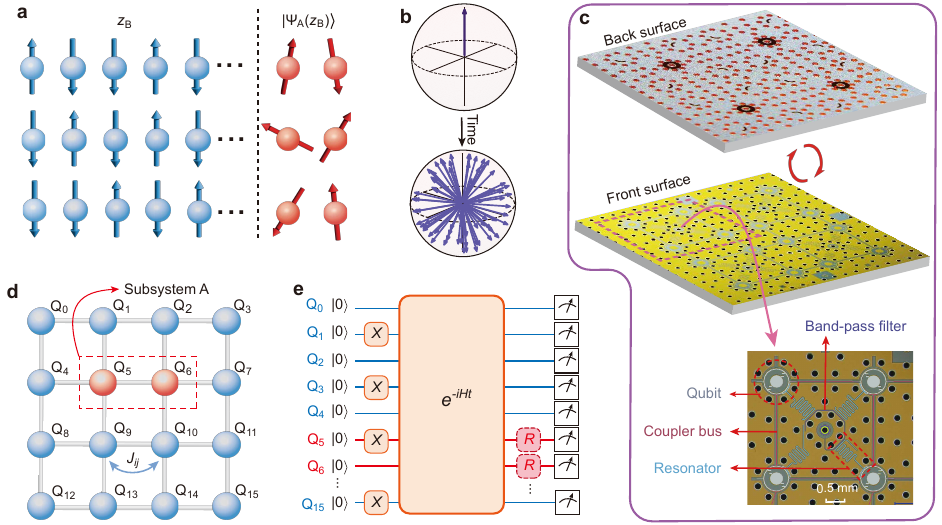}
	\caption{\textbf{Projected ensembles and experimental setup.}
		\textbf{a,} Schematic diagram of a projected ensemble. Red and blue spins represent subsystems A and B, respectively. Each projected result $z_\mathrm{B}$ corresponds to a measurement outcome $\ket{\Psi_\mathrm{A}(z_\mathrm{B})}$.  
		\textbf{b,} Sketch of deep thermalization. The system starts from a pure state, where the projected ensemble distributes on a local region of the Hilbert space. After a long-time evolution, the states of the projected ensemble distributes as a Haar ensemble, uniformly over the Hilbert space.
		\textbf{c,}~Optical images of the front and back sides of 16-qubit superconducting chip. The bottom false-colored image covers a single unit cell on the front side.  Each unit (pink dashed box) involves four qubits, with each qubit being coupled to a $\lambda/4$  readout resonator (light blue). These four readout resonators are then coupled to the bandpass filter (purple). Neighboring two qubits are coupled capacitively via a coupler bus (red). The black holes are superconducting TSVs. The TSV that locates at the center of each unit is connected to a readout port backside and is coupled capacitively to the  $\lambda/4$  band-pass filter at the front.
		\textbf{d,} Effective lattice model of the experimental system. There are 16 qubits arranged in a $4\times 4$ square lattice  with nearest-neighbor spin-exchange interaction. The subsystem $A$ contains two qubits: Q$_5$ and Q$_6$.
		\textbf{e,}~Experimental procedure. The initial state is prepared with $X$ gates on the target qubits.
		Then, all qubits are tuned into resonance, and the system evolves under the Hamiltonian $\hat H$.
		After a time $t$, we perform the joint single-shot readout on all qubits.
		The dashed boxes $R$ are identity, $X/2$, or $Y/2$ gate, to realize state tomography.
	}
	\label{fig_1}
\end{figure}

\begin{figure}[t] \includegraphics[width=0.5\textwidth]{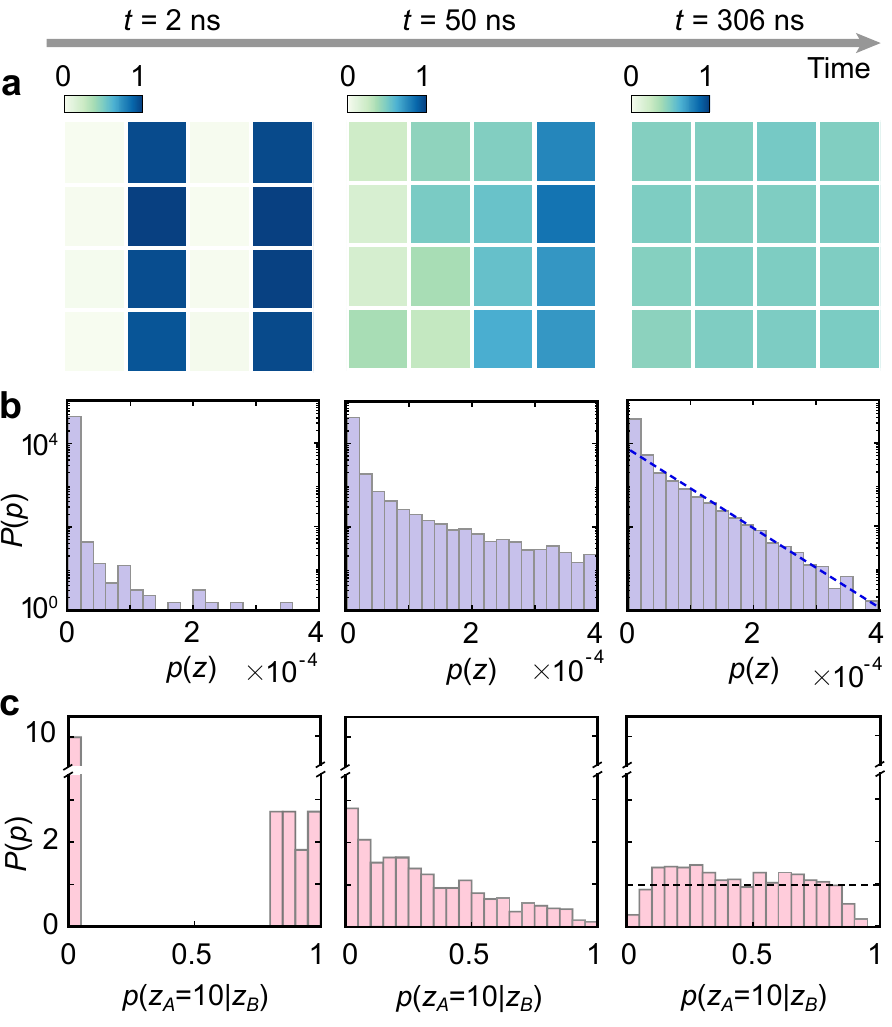}
	\caption{\textbf{Experimental signatures of ergodicity.} \textbf{a,} Distributions of the qubit excitations $n_j$ for evolution times  $t=2$~ns, $50$~ns, and $306$~ns.  \textbf{b,} Statistics of the bit-string probability $p(z)$ for the different evolution times. The blue dashed line denotes an exponential fitting $P(p)=\mathcal{D} e^{-\mathcal{D}p}$, representing the Porter-Thomas distribution.
		\textbf{c.} Distributions of the conditional probability $p(z_\mathrm{A}=10|z_\mathrm{B})$ for different evolution times. The black dashed horizontal line denotes the uniform distribution.
	}
	\label{fig_2}
\end{figure}

\begin{figure*}[t] \includegraphics[width=1\textwidth]{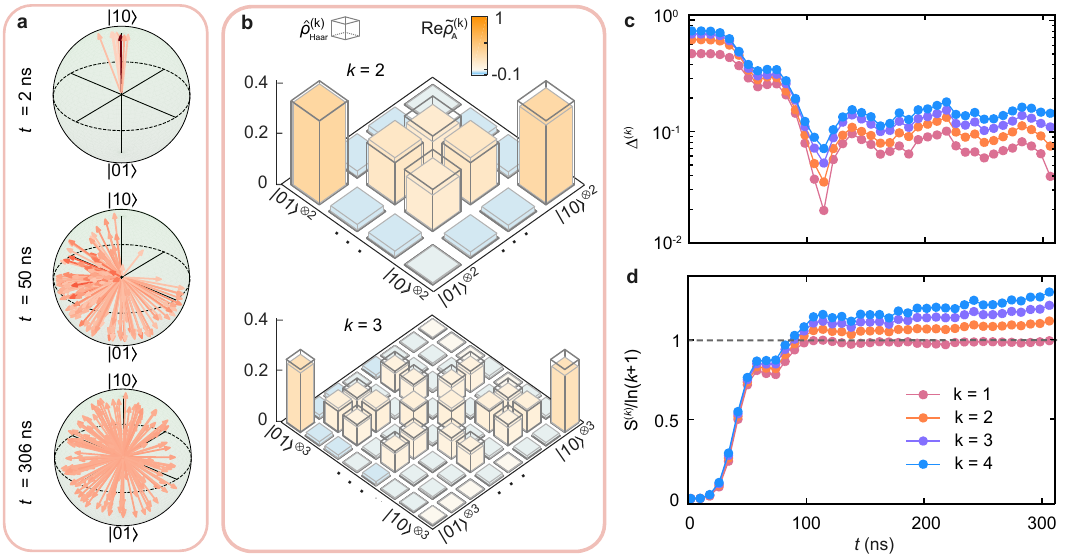}
	\caption{\textbf{Experimental signatures of deep thermalization.}
		\textbf{a.} Distributions of $\tilde{\rho}_{\mathrm{A}}(z_\mathrm{B})$ on the Bloch sphere for  $t=2$~ns, $50$~ns, and $306$~ns. The north and south pole denote the  states $\ket{10}$ and $\ket{01}$, respectively. 
		The intensity of arrows' color relates the probability $p(z_\mathrm{B})$.
		\textbf{b,} Real parts of the second- and third-moment density matrices of $\mathcal{E}^{\text{hf}}_{\Psi, \mathrm{A}}$ at $t=306$~ns.
		The empty boxes are the corresponding density matrices of the Haar ensemble.
		\textbf{c,} Dynamics of the trace distance $\Delta^{(k)}$ between the $k$th-moment density matrices~($k$=1,2,3,4) of $\mathcal{E}^{\text{hf}}_{\Psi, \mathrm{A}}$ and the Haar ensemble.
		\textbf{d,} Dynamics of $S_\mathrm{A}^{(k)}/\ln (k+1)$, which is the normalized entropy of $\tilde{ \rho}^{(k)}_{\mathrm{A}}$.  
	}
	\label{fig_3}
\end{figure*}

\begin{figure}[t] \includegraphics[width=0.5\textwidth]{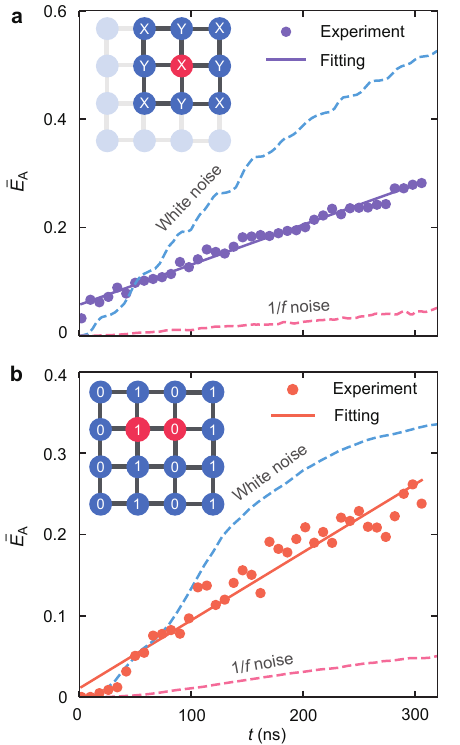}
	\caption{\textbf{Benchmark of many-body information leakage.}
		\textbf{a,}~Dynamics of the averaged entropy $\bar E_{\mathrm{A}}$ of the projected ensembles for a 9-qubit system. The inset shows the system and the initial state. The red and blue qubits are subsystems A and B, respectively, and $X$ ($Y$) corresponds to the initial states being the eigenstate of $\hat\sigma^{x}$ ($\hat\sigma^{y}$) with the positive eigenvalue. 
		\textbf{b,}~Dynamics of the averaged entropy $\bar E_{\mathrm{A}}$ with spin-conserved initial states for a 16-qubit system.
		We consider all charge-conserved sectors of  projected ensembles.
		The system structures and the initial states are also shown in the insets.
		The solid curves are  linear fits.
		The dashed curves are  numerical results with white and $1/f$ noise, respectively (Methods).
	}
	\label{fig_4}
\end{figure}

\begin{methods}
	
\section*{Experimental setup}

Our experiment is performed in a dilution refrigerator with a base temperature about $10$ mK at the mixing-chamber stage.
We use a scalable two-dimensional superconducting quantum processor, 
which contains 16 qubits arranged in a $4\times4$ array (Fig.~\ref{fig_1}c).
We implement a floating frequency-tunable transmon qubit with concentric and gradiometric geometry, optimized through the surface-participation-ratio analysis \cite{wang2015surface}. The loop size of the concentric qubit is approximately $3 \times 10^5$~\textmu m$^2$.

Each qubit enables full controls ($XY$ and $Z$),
and the control signals access the qubits via spring contacts, featuring vertical wiring. 
In addition, we achieve the $X$-crosstalk below $1 \times 10^{-3}$ and $Z$-crosstalk below $5 \times 10^{-3}$ between any control port $i$ and qubit Q$_j$ ($i \neq j$).
With the gradiometric geometry and optimized widths of the electrodes, the qubit $1/f$ flux noise is  about 4.9~$\mu\Phi_0/\sqrt{\mathrm{Hz}}$ at 1~Hz.

For readout, a band-pass filter \cite{jeffrey2014fast} is used to indirectly couple the readout resonators to the transmission line, achieving an average qubit--resonator coupling strength of approximately $2\pi \times 150$~MHz  while maintaining the Purcell limit for the qubit relaxation times longer than $1$~ms. This allows us to achieve a high signal-to-noise ratio (SNR) for the readout,
and the average simultaneous readout assignment fidelities are about $99.6\%$ and $97.5\%$ for the states $\ket{0}$ and $\ket{1}$, respectively.

The anharmonicity of the qubits ranges from $-248$ to $-218$~MHz, which is more than 40~times larger than the coupling strength between the neighboring qubits. Thus, our system is in the hard-core limit, which can be described by a 2D spin-$\frac{1}{2}$ XY model in Eq.~(\ref{H}).
More details about our experimental setup are provided in Supplementary Information.

\section*{Measuring projected ensembles}
To measure projected ensembles, joint measurements with single-shot readout of all qubits are required. 
Prior to readout, the measurement basis for subsystem A is adjusted by applying rotation gates  $R$  on the target qubits to perform quantum state tomography (Fig.~\ref{fig_1}e). For a two-qubit subsystem  A, quantum state tomography involves 9 measurement bases: 
$\{XX$, $XY$, $XZ$, $YX$, $YY$, $YZ$, $ZX$, $ZY$, $ZZ\}$. For each basis $v_\mathrm{A}$, we collect  $M(v_\mathrm{A})$  single-shot outcomes, represented as bit strings of 16 qubits, with the count of each bit string  $z$  denoted as  $m_z$.
Our system achieves high readout fidelities and low readout crosstalks. To further mitigate readout errors, we optimize the measurement results using an approximate readout transition matrix~\cite{PhysRevLett.118.210504}:
\begin{align} \label{Ea}
	\mathbb{F} = \otimes_{j=1}^{N} \begin{bmatrix}
		F_{00}^j  & 1-F_{11}^j \\
		1-F_{00}^j  & F_{11}^j\\
	\end{bmatrix},
\end{align}
where  $N$ is the number of qubits, and  $F_{00(11)}^j$  represents the readout fidelity of state \(\ket{0}\)(\(\ket{1}\)) for Q$_j$. 
Using this matrix, the optimized count of measurements for a given bit string  $z$  is calculated as
\begin{align} \label{Ea}
	\tilde{m}_z = \sum_{z'} \mathbb{F}_{z, z'} m_{z'}.
\end{align}

Then, we determine the optimized number of measurements for each bit string  $z_\mathrm{B}$  under a given basis  $v_\mathrm{A}$, labeled as $\tilde{m}_{z_\mathrm{B}}(v_\mathrm{A})$. Only bit strings  $z_\mathrm{B}$  satisfying $\tilde{m}_{z_\mathrm{B}}(v_\mathrm{A}) \geq 80$ for all bases  $v_\mathrm{A}$  are considered. From the single-shot results, we estimate the corresponding density matrix $\hat \rho_\mathrm{A}(z_\mathrm{B})$. The probability of measuring  $z_\mathrm{B}$  is then calculated as
\begin{align} \label{Ea}
	p(z_\mathrm{B}) = \frac{\sum_{v_\mathrm{A}} \tilde{m}_{z_\mathrm{B}}(v_\mathrm{A})}{\sum_{v_\mathrm{A}} M(v_\mathrm{A})}.
\end{align}
This procedure allows for the reconstruction of the projected ensemble.

For the 9-qubit experiment, the number of readouts per basis,  $M(v_\mathrm{A}) $, is approximately  $8 \times 10^4$.
For the 16-qubit experiments (with charge-conserved initial states),  $M(v_\mathrm{A})$  varies with the evolution time, increasing as the evolution progresses, where $M(v_\mathrm{A})$  ranges from  $2 \times 10^5$  to  $5 \times 10^5$.

\section*{Benchmarking many-body information leakage}

In superconducting circuits, dephasing can be modeled by a time-dependent Hamiltonian:
\begin{align} \label{Ht}
	&\hat H(t) = \hat H + 
    \frac{1}{2}\sum_{j} \xi_j(t)\hat\sigma^z_j,\\
   & \braket{\xi_j(0)\xi_j(t)} = \int_{-\infty}^\infty \frac{d\omega}{2\pi}S_j(\omega)e^{-i\omega t}.
\end{align}
where $\xi_j(t)$ represents Gaussian fluctuations, and $S_j(\omega)$ denotes the noise spectral density.
For white and $1/f$ noise, the spectral densities satisfy $S_j(\omega) = W_j$ and $S_j(\omega) = A_j/|\omega|$, respectively.
In the numerical simulations, we first extract the noise intensities $W_j$ and $A_j$ from the experimentally measured $T^*_2$ values at the operating point (see Fig.~S5 in the Supplementary Information). 
In the experiment, each measurement cycle takes approximately 600~\textmu s, and the total time required to measure a projected ensemble is $10^3$-$10^4$~s.
Thus, for the $1/f$ noise, the low- and high-frequency cutoffs are set to 1 mHz and 100 kHz, respectively. 
For the white noise, a high-frequency cutoff of 1 GHz is used.
We then compute $\bar{E}_\mathrm{A}$ by solving the Schrödinger equation governed by the Hamiltonian in Eq.~(\ref{Ht}). All simulations are performed using the QuTiP package~\cite{JOHANSSON20131234}.

\end{methods}

\noindent{\bfseries References}\setlength{\parskip}{12pt}%

	\providecommand{\noopsort}[1]{}\providecommand{\singleletter}[1]{#1}%



\begin{addendum}
	\item [Acknowledgments] Z.Y. and Y.N. are partially supported by Ministry of Education, Culture, Sports, Science and Technology (MEXT) Quantum Leap Flagship Program (QLEAP)
    (via Grant No. JPMXS0118068682). F.N. is supported in part by:
      the Japan Science and Technology Agency (JST)
       [via the CREST Quantum Frontiers program Grant No. JPMJCR24I2,
      the Quantum Leap Flagship Program (Q-LEAP), and the Moonshot R\&D Grant Number JPMJMS2061],
      and the Office of Naval Research (ONR) Global (via Grant No. N62909-23-1-2074).
    Y.R.Z. is supported in part by: the National Natural Science Foundation of China (via Grant No. 12475017), the Natural Science Foundation of Guangdong Province (via Grant No. 2024A1515010398).
	
	\item[Author contributions] Z.Y., Z.-Y.G., F.N., and Y.N. conceived the idea and experiment. 
    Z.Y. performed the experiments.
    Z.-Y.G contributed to the underlying theory.
    Z.Y. improved the spring contacts for DC-bias, 
        designed and fabricated the chip, and set up the measurement system. R.L. and Z.Y. developed the measurement code. 
	Z.-Y.G and Z.Y. performed data analysis and wrote the manuscript.
	All authors contributed to revising the manuscript and Supplementary Information. 
	F.N. and Y.N. supervised this project.
	
	\item[Competing Interests] The authors declare no competing interests.
	
\end{addendum}


\clearpage 


\section*{Supplementary material}

## Appendix A: Experimental wiring setup

The measurement setup for our superconducting qubit system is illustrated in Fig. S1. The sample is housed within a Blufors LD400 dilution refrigerator, which maintains a base temperature of approximately 10 mK at the mixing-chamber (MC) stage. For each unit consisting of four qubits and one readout port, we employ a near-quantum-limited impedance-matched parametric amplifier (IMPA) [1] to amplify the readout signals at the MC stage, achieving an amplitude gain of around 20 dB with a 600–700 MHz bandwidth.

A DC source (Yokogawa GS200) is utilized to apply the DC coil current to flux-bias the IMPA at its operation frequency. As shown in Fig. S1, twisted-pair DC cables are used in the flux-bias line to mitigate the electric and magnetic noise from the environment. In addition, to suppress the high-frequency noise, a $\pi$-filter (cutoff ~ 1 MHz) at room temperature and a homemade RC filter (cutoff ~ 2 kHz) at the 4 K plate are implemented. In the setup, a frequency doubler (Marki ADA-1030) is used to generate the high-frequency pump signal for the IMPA, which ranges from 16.2 to 16.9 GHz, approximately at twice the readout-resonator frequency. The long-time readout results shown in Fig. S15 indicate that the DC source and the frequency doubler remained stable during the experiments.

In our setup, the quantum analyzer (Zurich Instruments UHFQA) is used to generate the readout input (RIN) signals and demodulate the readout output (RO) signals. As shown in Fig. S3b, four readout resonators are coupled to a single readout port in one unit on our chip, allowing us to perform frequency-multiplexed readout through the four readout resonators. To ensure that the local-oscillator (LO) signals for the modulation and demodulation are phase-locked, we employ a Keysight microwave source to generate the LO signal, which is then amplified and goes through a power splitter to produce two phase-locked LO signals: one for the modulation and the other for the demodulation. The Zurich Instruments HDAWG8, an arbitrary waveform generator (AWG), generates all $XY$ and $Z$ control signals for the tunable transmon qubits [2]. As depicted in Fig. S1, we utilize two AWG ports to generate the in-phase (I) and quadrature (Q) signals, which are then multiplexed with a LO signal by an IQ Mixer (Marki MMIQ-0218L) to produce the drive signal for the qubit. The HDAWG8 supports DC settings, allowing us to generate the DC and fast $Z$-pulse from a single output port. We combine the $XY$ and $Z$ signals at room temperature through a diplexer (QMC-CRYODPLX-0218). Then, the combined signals are sent to the qubit through microwave cables in the refrigerator. Using this approach, we can fully control each qubit with just one microwave

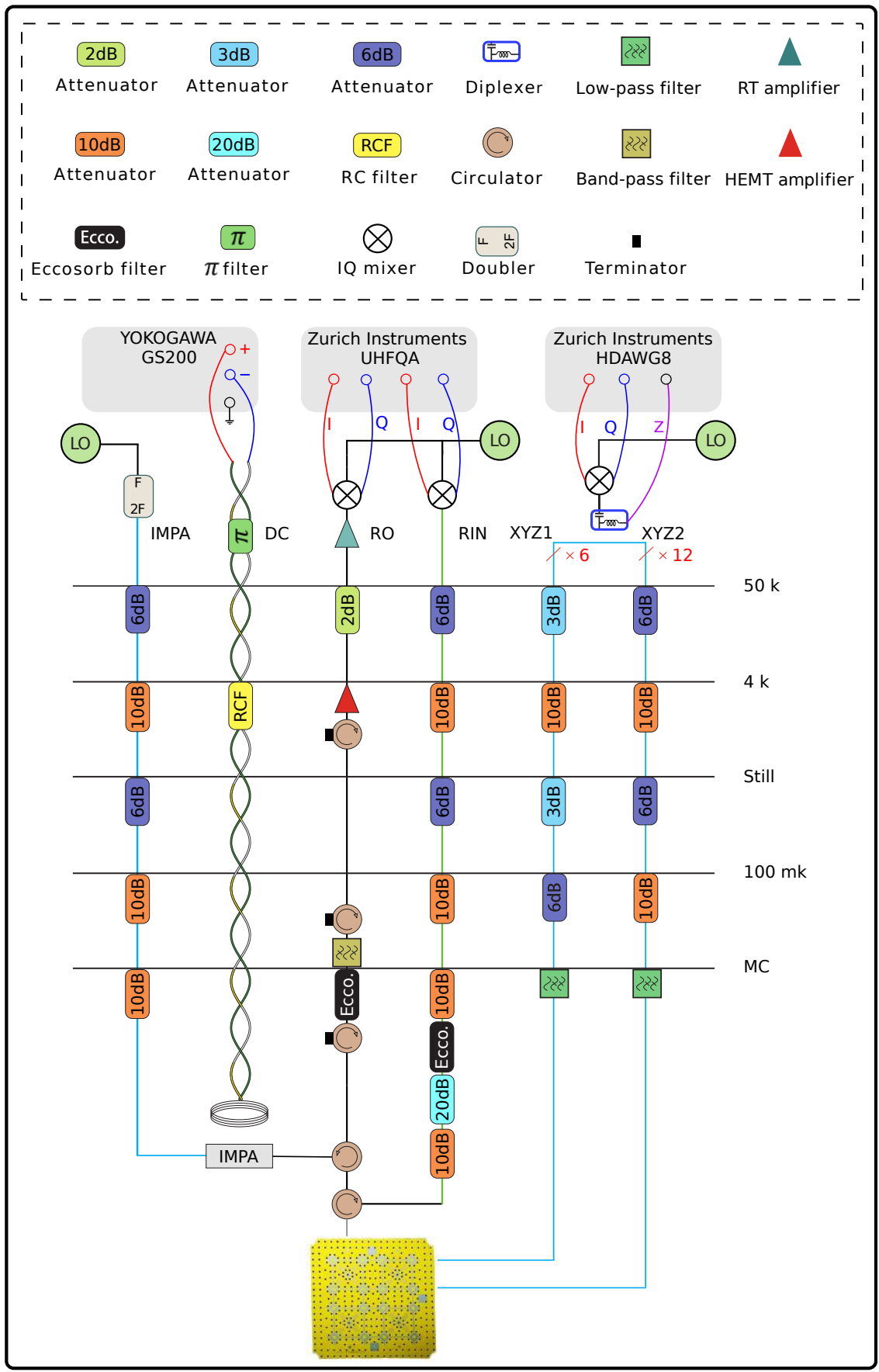

FIG. S1. **Measurement setup**.

coaxial cable.

In the experiment, we require both low-frequency and high-frequency signals, which are used respectively to modulate the transmon qubit frequency and to drive the qubit transition. For the DC signal, we do not want it to be attenuated at the MC stage to avoid generating heat which might be transferred to the qubit and/or lead to a temperature increase of the MC plate. On the other hand, for the high-frequency drive signal, it is essential to be attenuated at the MC stage to suppress the high-frequency noise coming from the high-temperature plates. To address this issue, we developed a special low-pass filter. As shown in Fig. S2, it features a 2 GHz passband for the low-frequency flux-bias signal and a relatively flat passband (4.5–6 GHz) with an attenuation of approximately 10 dB for the high-frequency drive signal. In developing this low-pass filter, three key considerations were taken into account:

1. **Sufficient attenuation at the qubit frequency**: The filter must provide enough attenuation at the qubit frequency to suppress the qubit thermal excitations.

2. **Flat response in the frequency domain**: It is essential to maintain a flat frequency response within the range of the qubit frequency. It is critical for the waveforms of the qubit drive signal, particularly for the short-time pumping signals. As shorter signals have larger bandwidths, excessive variations in attenuation across the frequency range would significantly distort the signals in the time domain.

3. **Thermal contact to the MC plate**: Ensuring good thermal contact between the low-pass filter and the MC plate. Poor thermal contact could result in the filter's temperature exceeding that of the MC plate, allowing high-temperature noise to reach the qubit.

In our wiring setup, we maintained two different wiring configurations for qubit control: the drive lines for $Q_3$, $Q_7$, $Q_8$, $Q_9$, $Q_{12}$ and $Q_{13}$ are designated as XYZ1 and the other lines are designated as XYZ2. The qubits connected to the drive lines XYZ1, even with the total attenuation 10 dB less than that in the drive lines XYZ2, showed comparable thermal population and $T_1$ (see Table I and Figs. S5d and S5e). The observed variations in the qubit thermal population may arise from the attenuation variations in the low-pass filter. In particular, $Q_9$ with the XYZ1 wiring configuration and the total attenuation of only 22 dB exhibits a thermal population of 0.66% and $T_1 \approx 50$ µs at the frequency of 5715 MHz.

## Appendix B: Device design

Our experiments are conducted on a scalable, two-dimensional superconducting quantum processor and enables full qubit control ($Z$- and $XY$-control). All the control signals access the quantum processor via

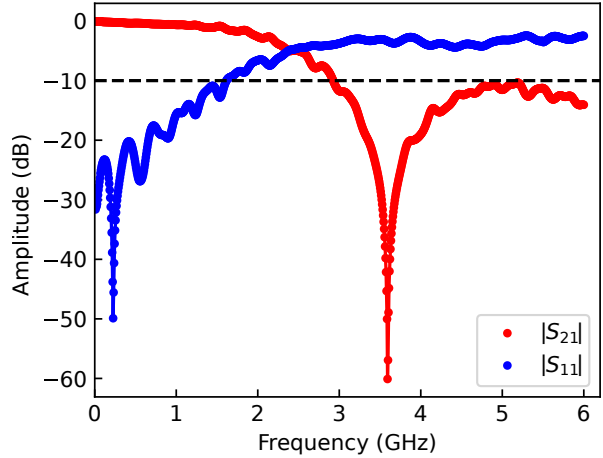

FIG. S2. $|S_{11}|$ and $|S_{21}|$ of the homemade low-pass filter.

spring contacts vertical coaxial-cable wiring. As depicted in Fig. S3a, the $4 \times 4$ tunable transmon qubit chip is mounted onto a stage featuring a 2D spring-contact array, aligned with the corner pillars. Figure S3d presents the cross-section along the dashed line in Fig. S3a. All the spring contacts directly contact the backside of the chip. The $Z/XY$-control pin is connected to the back end of the qubit, with the contact point (green) shown in Fig. S3c. Each readout pin is connected to a readout superconducting through-silicon via (TSV; yellow-green), which couples capacitively to the $\lambda/4$ band-pass filter (purple) on the top side of the chip, functioning as a transmission line.

As shown in Fig. S3b, each $\lambda/4$ band-pass filter couples to four readout resonators (light blue), allowing frequency-multiplexed readout for each unit consisting of four qubits. In addition, numerous TSVs are placed in the chip, connecting the segmented topside grounds to the backside ground to suppress potential slot-line or substrate modes. Moreover, these TSVs help to localize the control signals, reducing unwanted crosstalk. In our case, we achieved $X$-crosstalk below $1 \times 10^{-3}$ and $Z$-crosstalk below $5 \times 10^{-3}$ between any control port $i$ and qubit $Q_j$ ($i \neq j$).

As shown in Fig. S3b, we implement floating tunable transmon qubits (light gray) with concentric geometry optimized through the surface-participation-ratio analysis [3]. The loop size of our concentric qubit is approximately $3 \times 10^5$ $\mu m^2$, nearly three orders of magnitude larger than typical SQUID loops used for qubits. However, as illustrated in Fig. S4b, with the gradiometric geometry and optimized linewidth in the qubit design, the square-root amplitude $\sqrt{A_\Phi}$ of the $1/f$ flux noise spectral density $A_\Phi/\omega$ is about 4.9 $\mu\Phi_0$, which is comparable to those in conventional designs [4].

To couple adjacent qubits, a coupler bus (dark red) is employed. The Hamiltonian for two adjacent

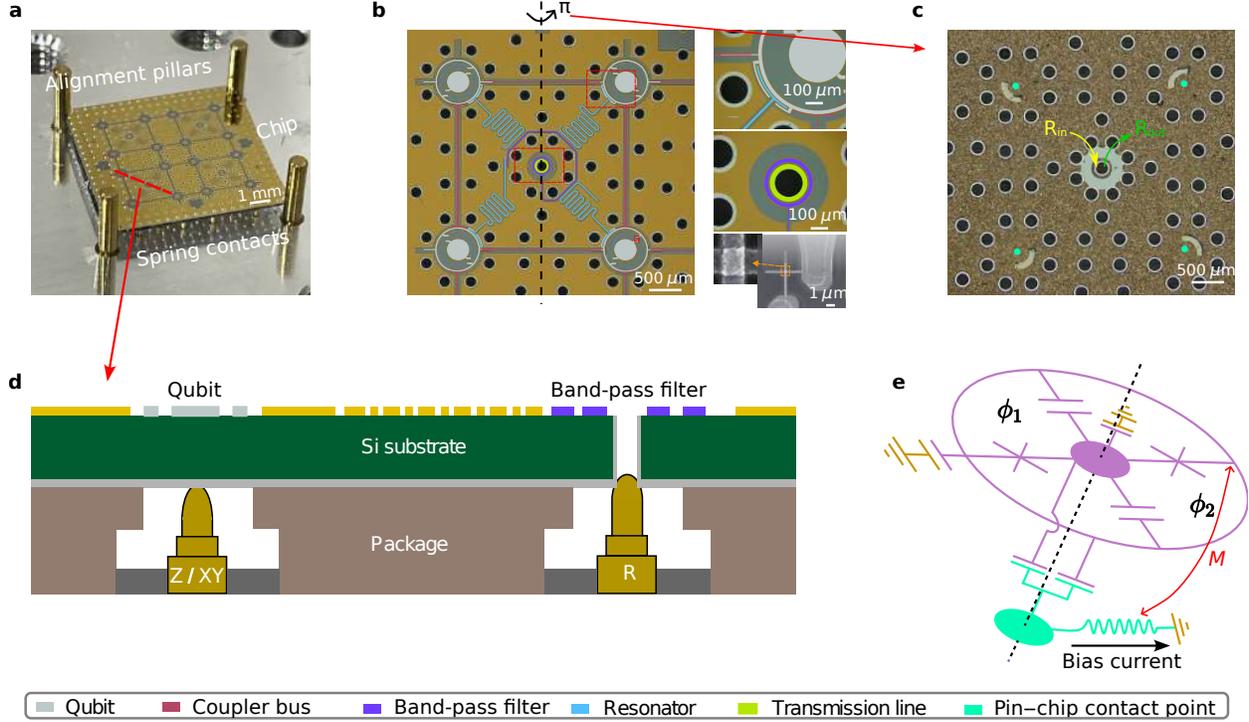

FIG. S3. **Package and the chip. a**, Optical images of the assembly of the package and chip. **b**, False-colored image of the top side of the chip. The insets show the magnified images of the parts enclosed in the red squares. The Josephson junction area is approximately 180×210 nm². **c**, Backside of the chip. **d**, Schematic diagram of the cross-sectional profile along the dashed line in **a**. **e**, Schematic circuit diagram of the qubit and control circuits.

qubits, $Q_i$ and $Q_j$, along with the coupler bus, is given by

$$\hat{H}/\hbar = \omega_i \hat{\sigma}_i^+ \hat{\sigma}_i^- + \omega_j \hat{\sigma}_j^+ \hat{\sigma}_j^- + \frac{1}{2}\omega_c \hat{a}^\dagger \hat{a} + g_j(\hat{\sigma}_j^+ \hat{a} + \hat{\sigma}_j^- \hat{a}^\dagger) + g_i(\hat{\sigma}_i^+ \hat{a} + \hat{\sigma}_i^- \hat{a}^\dagger) + g_{ij}(\hat{\sigma}_i^+ \hat{\sigma}_j^- + \hat{\sigma}_i^- \hat{\sigma}_j^+). \quad \text{(S1)}$$

Here, $\omega_i$, $\omega_j$, and $\omega_c$ denote the frequencies of $Q_i$, $Q_j$ and the coupler, respectively; $g_i$ ($g_j$) represents the coupling strength between $Q_i$ ($Q_j$) and the coupler. The parameter $g_{ij}$ denotes the direct coupling strength between two nearest-neighbor qubits. The operators $\hat{\sigma}^\alpha$ ($\alpha = x, y, z$ and $\hat{\sigma}^\pm = \hat{\sigma}_j^x \pm \hat{\sigma}_j^y$) correspond to the Pauli matrices of qubits, while $\hat{a}$ and $\hat{a}^\dagger$ represent the lowering and raising operators of the coupler mode. By applying the Schrieffer–Wolff transformation [5],

$$\hat{U} = \exp\left[\sum_{\lambda=i,j} \frac{g_\lambda}{\Delta_\lambda}\left(\hat{\sigma}_\lambda^+ \hat{a} + \hat{\sigma}_\lambda^- \hat{a}^\dagger\right)\right]. \quad \text{(S2)}$$

The coupler can be effectively decoupled from the system up to second order in $g_\lambda/\Delta_\lambda$, yielding an effective XY model

$$\hat{H}_{\text{eff}}/\hbar = \sum_{\lambda=i,j} \tilde{\omega}_\lambda \hat{\sigma}_\lambda^+ \hat{\sigma}_\lambda^- + \left(\frac{g_i g_j}{\Delta} + g_{ij}\right)\left(\hat{\sigma}_i^+ \hat{\sigma}_j^- + \hat{\sigma}_i^- \hat{\sigma}_j^+\right). \quad \text{(S3)}$$

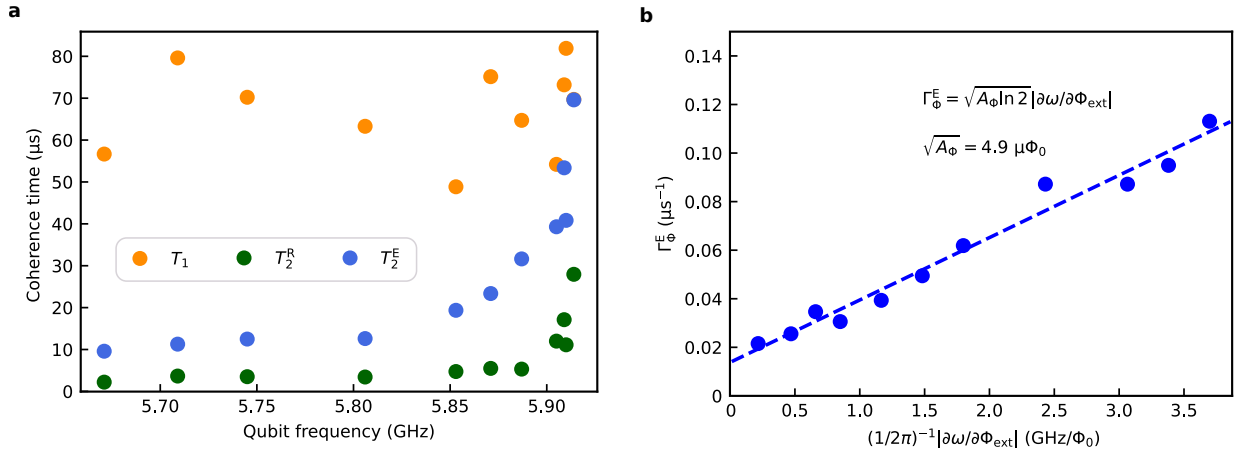

FIG. S4. **Decoherence characterization. a**, Qubit coherence time $T_1$ (energy relaxtion), $T_2^{\mathrm{R}}$ (Ramsey), and $T_2^{\mathrm{E}}$ (spin-echo) as a function of the qubit frequencies. The data is for Q$_{11}$. **b**, Pure dephasing rates $\Gamma_\varphi^{\mathrm{E}}$ as a function of the slope of the qubit spectrum. We derive the $1/f$ flux-noise amplitude $\sqrt{A_\Phi}$ through the linear fit to the data.

Here, the coupler is assumed to remain in its ground state, which holds true in our experiment. $\Delta_\lambda = \omega_\lambda - \omega_c$ is the frequency detuning between qubit Q$_\lambda$ and the coupler. $\Delta$ is the effective detuning given by $1/\Delta = \left(1/\Delta_i + 1/\Delta_j\right)/2$. $\tilde{\omega}_\lambda = \omega_\lambda + \left(g_\lambda^2/\Delta_\lambda\right)\left(\alpha_\lambda/\left(\Delta_\lambda + \alpha_\lambda\right)\right)$ is the Lamb-shifted qubit frequency, where $\alpha_\lambda$ represents the anharmonicity of Q$_\lambda$. In our device, the frequency of the coupler buses used to connect adjacent qubits within a unit is about 22 GHz, while the frequency of the coupler buses used to connect adjacent qubits between neighboring units is approximately 40 GHz. The considerable detuning between the qubit and the coupler bus can effectively suppress the unwanted qubit–coupler interactions to the residual coupling strength below 20 kHz. On the other hand, the effective coupling strength between adjacent qubits is approximately 3.5 MHz within one unit and 4.5 MHz between neighboring units, as indicated in Fig. S6f.

In our design, we carefully optimized the frequencies of adjacent coupler buses to maintain a frequency detuning exceeding 1 GHz, thereby effectively mitigating unwanted interaction between the next-nearest-neighbor (NNN) qubits through the virtual exchange interaction, which, in our setup, is very small (∼ 0.08 MHz) and can therefore be considered negligible. The magnified micrographs in the insets of Fig. S3b illustrate the following regions: the coupling capacitor between the transmon qubit and the readout resonator, the capacitor between the transmission line and the band-pass filter, and the Manhattan-style superconducting Josephson junction [6].

As illustrated in Fig. S3c, the control-line coupler of each qubit is patterned on the backside of the chip with a minimalistic design, consisting solely of an arc-shaped gap around the pin–chip contact point (green). In Fig. S3e, the schematic of the qubit and control circuits is presented, illustrating how the microwave-

frequency drive signals and low-frequency bias signals couple to the qubit. When a DC signal is applied to the control pin, the current is asymmetrically distributed around the gap, generating a differential magnetic flux $\Phi = \Phi_1 - \Phi_2$ within the superconducting qubit loop, where $\Phi_1$ and $\Phi_2$ denote the respective magnetic fluxes penetrating through the two half SQUID loops. This differential magnetic flux $\Phi_{\mathrm{ext}}$ is the effective magnetic flux bias for the tunable transmon qubit. The dependence of the qubit frequency on $\Phi_{\mathrm{ext}}$ can be approximated by [2]:

$$\omega(\Phi_{\mathrm{ext}}) = \left(\omega_{\mathrm{max}} + \frac{E_C}{\hbar}\right)\sqrt{d^2 + (1 - d^2)\cos^2\left(\frac{\pi\Phi_{\mathrm{ext}}}{\Phi_0}\right)} - \frac{E_C}{\hbar}. \tag{S4}$$

Here, $E_C$ is the transmon charging energy, and $\omega_{\mathrm{max}} = (\sqrt{8E_J E_C} - E_C)/\hbar$ represents the maximum qubit frequency. The asymmetry parameter $d$ of the SQUID junctions is defined as $d = |(E_{J,2} - E_{J,1})/(E_{J,2} + E_{J,1})|$, where $E_{J,1}$ and $E_{J,2}$ denote the Josephson energies of the two SQUID junctions. It is important to note that Eq. (S4) is an approximation, valid only in the regime where $E_J \gg E_C$. When a high-frequency drive signal is applied, the electromagnetic field couples to the qubit via the effective mutual capacitance. In our chip, the control port exhibits mutual inductive ($Z$) coupling of approximately 0.8 pH. The capacitive ($XY$) coupling to the qubit is represented in the qubit decay rate of $2\pi \times 200$ Hz to the port. The inductive and capacitive coupling strengths can be independently modulated through modifications to the angle and size of the gap in the backside ground plane.

## Appendix C: Fabrication

Our sample is fabricated on a 3-inch silicon wafer with (100) crystal orientation and a resistivity of more than $20\,\mathrm{k\Omega \cdot cm}$. The fabrication process is outlined as follows:

1. **Through-silicon vias (TSVs) fabrication**: Etch the vias using the deep reactive ion etching technique.

2. **TiN-film deposition**: Deposit 100-nm-thick TiN films on both sides of the silicon wafer with sputtering at a temperature of 850°C immediately after the wafer pre-cleaning.

3. **Via metallization**: Evaporate 300-nm-thick aluminum with an incident angle of 40° on both the front and back sides of the wafer on a planetary rotating stage.

4. **Aluminum-film etching**: Laminate resist films on both sides of the wafer and perform UV exposure and development, leaving the resist covering the vias. Perform wet etching to remove the aluminum on both sides of the wafer.

5. **TiN-film etching**: Spin-coat the wafer with AZ1500 photoresist, Then, do the photolithography using direct laser writing on each side of the wafer sequentially to define the qubits and resonators on the top surface and control-line contacts on the backside. Subsequently, perform dry etching on both sides after development.

6. **Cleaning**: Immerse the wafer in Remover104 at 120°C for more than four hours, then perform the oxygen plasma cleaning and HF cleaning.

7. **Junction fabrication**: Spin-coat the wafer with Copolymer and PMMA sequentially to form a bilayer electron-beam resist structure. Then, expose the wafer in an electron-beam writer. After developing the resist, evaporate aluminum films for the junctions (Fig. S3b) in an e-gun evaporator.

8. **Indium evaporation**: Cover the backside of the wafer with lamination film resist. Then, expose the wafer by a direct laser writer. After the development, evaporate $4$-$\mu$m-thick indium on the backside of the chip.

9. **Dicing and lift-off**: Dice the wafer into chips and finally perform the lift-off.

## Appendix D: Device parameters

Our superconducting quantum processor employs frequency-tunable transmon qubits with a gradiometric structure, which enhances resilience against uniform global flux noise. The maximum frequencies of the qubits alternate between high and low values, as depicted in Fig. S5b. This frequency arrangement effectively reduces the interactions among the adjacent qubits when we characterize them at the beginning without bias currents (all the qubits will be at their sweet points if there are no flux offsets). In the experiments (see Fig. S9), we use DC currents to bias all the qubits at the same operation frequency while preparing their initial states and performing their readout at their individual idle frequencies with additional $Z$-pulses applied. As demonstrated in Fig. S6e, by ensuring sufficient frequency detuning between all the neighboring qubits, we suppress the $ZZ$ coupling [7] to be below 110 kHz. This allows the fidelities of simultaneous single-qubit gates to be comparable to those of individual single-qubit gates. Moreover, the single-qubit gates are characterized by randomized benchmarking [8–10] at the idle frequencies with the $Z$-pulses applied. The achieved fidelities of the single-qubit gates suggest that both the $Z$-pulse distortion and crosstalk are well calibrated (see for the details in Sections E 1 and E 3.)

In our chip, the qubit anharmonicity $\eta$ (representing the nonlinear boson–boson interaction on site) ranges from $-248$ to $-218$ MHz, which is more than 40 times larger than the coupling strength between the

TABLE I. **Qubit parameters.** For better comparison, we show the simultaneous-readout assignment fidelity $F_{ii}$ (the probability of correctly classifying the qubit state $|i\rangle$ as $i$), qubit idle-point thermal population $P_{\text{th}}^{\text{idle}}$, qubit operation-point thermal population $P_{\text{th}}^{\text{op}}$ measured when all the qubits are aligned at the operation frequency, and the fidelities of individual $P_{\text{RB}}^{\text{ind}}$ and simultaneous $P_{\text{RB}}^{\text{sim}}$ single-qubit gates, performed at the idle points with the $Z$-pulses applied. RB stands for randomized benchmarking.

| Parameters | $Q_0$ | $Q_1$ | $Q_2$ | $Q_3$ | $Q_4$ | $Q_5$ | $Q_6$ | $Q_7$ |
|---|---|---|---|---|---|---|---|---|
| $F_{00}(\%)$ | 99.74 | 99.42 | 99.82 | 99.74 | 99.79 | 99.56 | 99.74 | 99.72 |
| $F_{11}(\%)$ | 98.75 | 98.82 | 98.85 | 98.24 | 99.05 | 98.43 | 98.61 | 97.09 |
| $P_{\text{th}}^{\text{idle}}(\%)$ | 1.39 | 1.9 | 2.58 | 0.75 | 0.46 | 3.81 | 5.39 | 2.46 |
| $P_{\text{th}}^{\text{op}}(\%)$ | 4.62 | 4.26 | 4.22 | 5.09 | 5.09 | 5.21 | 5.16 | 5.24 |
| $P_{\text{RB}}^{\text{ind}}(\%)$ | 99.25 | 99.44 | 99.48 | 99.66 | 99.54 | 99.52 | 99.05 | 99.44 |
| $P_{\text{RB}}^{\text{sim}}(\%)$ | 99.39 | 98.44 | 99.60 | 99.29 | 99.68 | 99.31 | 98.92 | 99.66 |

| Parameters | $Q_8$ | $Q_9$ | $Q_{10}$ | $Q_{11}$ | $Q_{12}$ | $Q_{13}$ | $Q_{14}$ | $Q_{15}$ |
|---|---|---|---|---|---|---|---|---|
| $F_{00}(\%)$ | 99.08 | 99.05 | 99.90 | 99.49 | 99.61 | 99.87 | 99.07 | 99.62 |
| $F_{11}(\%)$ | 96.68 | 98.87 | 96.67 | 97.82 | 98.85 | 98.57 | 97.55 | 97.20 |
| $P_{\text{th}}^{\text{idle}}(\%)$ | 2.85 | 2.6 | 1.22 | 0.56 | 4.04 | 4.12 | 4.13 | 2.63 |
| $P_{\text{th}}^{\text{op}}(\%)$ | 5.14 | 5.04 | 4.88 | 5.24 | 4.78 | 3.94 | 4.43 | 4.93 |
| $P_{\text{RB}}^{\text{ind}}(\%)$ | 97.91 | 99.44 | 99.44 | 99.58 | 99.81 | 99.43 | 99.63 | 99.61 |
| $P_{\text{RB}}^{\text{sim}}(\%)$ | 97.14 | 99.47 | 98.84 | 99.64 | 98.92 | 99.18 | 99.25 | 99.63 |

neighboring qubits. Consequently, our system is in the hard-core limit of the interacting boson model and well described by the XY model [11]. For the qubit-state readout, we use a resonator dispersively coupled to the transmon qubit to perform non-destructive measurement [2]. A band-pass filter [12, 13] is used to indirectly couple the readout resonators to the transmission line (as illustrated in Fig. S3), achieving an average qubit–resonator coupling strength of approximately $2\pi \times 150$ MHz (see Fig. S6b), while maintaining the Purcell-limited relaxation times for all the qubits longer than 1 ms. This results in a relatively large ratio between the dispersive shift $\chi$ and the readout-resonator decay rate $\kappa$ with the median value of $|2\chi/\kappa|$ being 0.7 across all the qubits, allowing us to attain a high signal-to-noise ratio (SNR) for the readout [14].

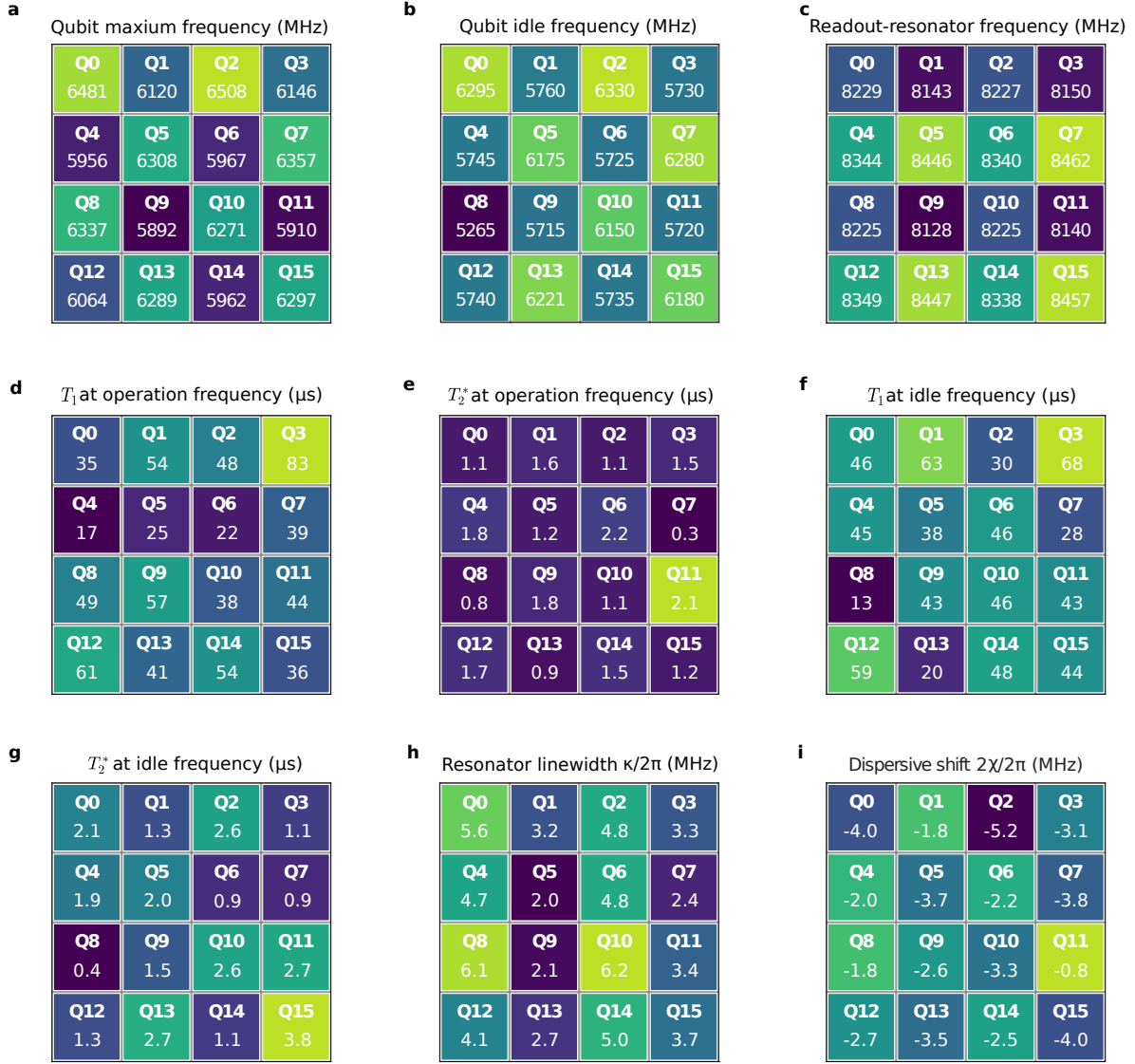

FIG. S5. **Chip parameters**. **a**, Qubit maximum frequencies. **b**, Qubit idle frequencies. **c**, Readout-resonator frequencies. **d** and **e**, $T_1$ and $T_2^*$ at the operation frequency. **f** and **g**, $T_1$ and $T_2^*$ at the idle frequencies. **h**, Readout-resonator linewidths $\kappa$. **i**, Dispersive shifts of the readout resonators, $2\chi$.

As illustrated in Fig. S7, the single-qubit gates and readout are performed at the idle points with the $Z$-pulses applied. During the waiting time of 600 μs, the $Z$-pulses are turned off. Then all the qubits are aligned to the operation point. With well-suppressed $Z$-pulse distortion and crosstalk, high-fidelity simultaneous single-qubit gates, and low readout crosstalk, we achieved an average simultaneous-readout assignment fidelity of 99.6% for state $|0\rangle$ and 97.5% for state $|1\rangle$ across all the qubits. Due to the extended waiting time at the operation point between two successive measurements, similar thermal populations $P_{\mathrm{th}}^{\mathrm{op}}$ were observed among all the qubits. Here, the thermal population of each qubit was estimated from

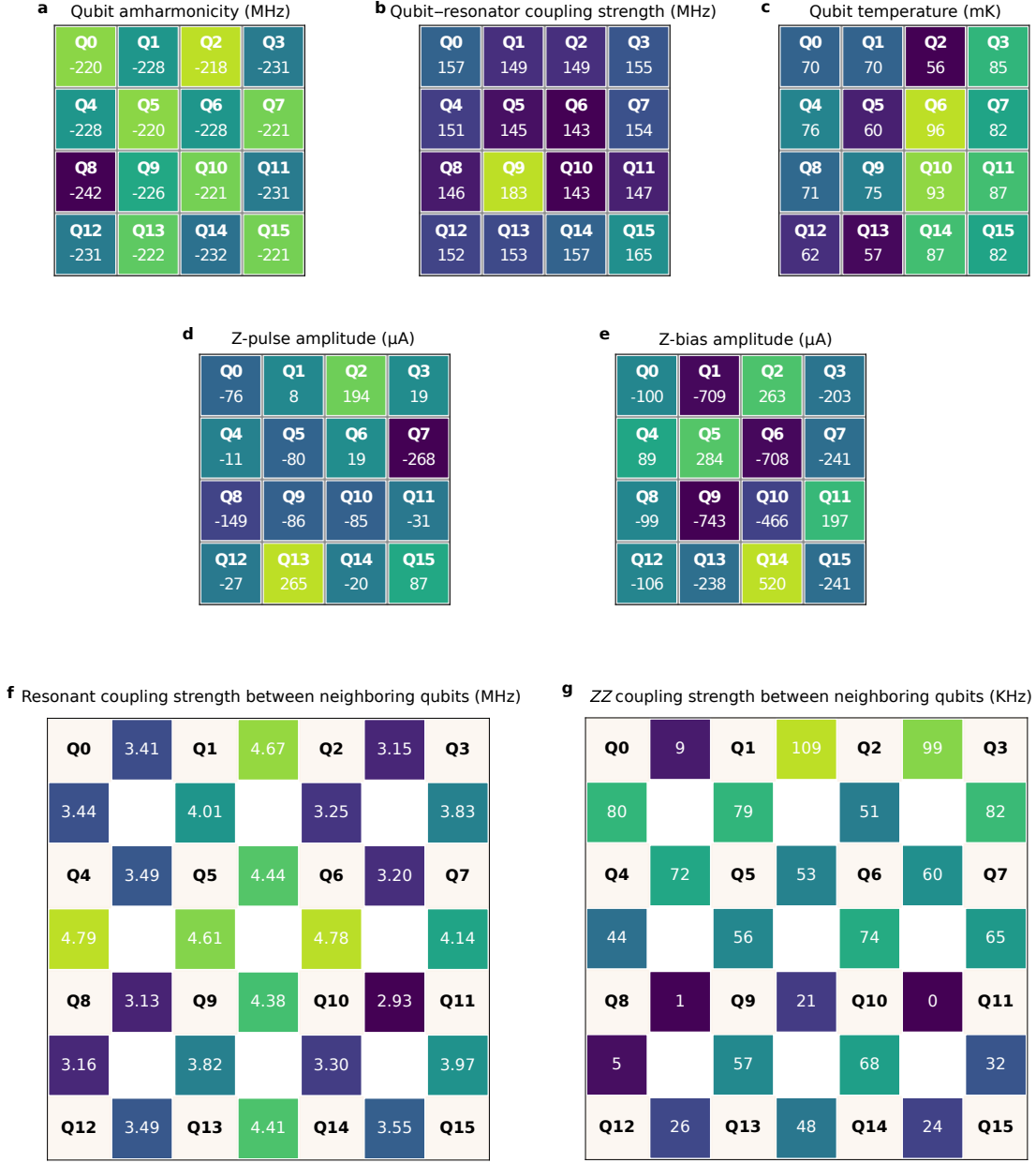

FIG. S6. **Chip parameters.** **a,** Qubit anharmonicity $\eta$. **b,** Qubit–resonator coupling strength. **c,** Effective qubit temperature at the idle point. **d,** $Z$-pulse amplitude used to bias the qubit from the working point to the idle point. **e,** $Z$-bias amplitude used to bias the qubit at the operation point. **f,** Coupling strength between the neighboring qubits. **g,** $ZZ$-coupling strength between the neighboring qubits.

the single-shot readout result from the first measurement $M_0$ and the second measurement $M_1$, given by $P_{\mathrm{th}} = F_{00}^0 - F_{00}^1$, where $F_{00}^1$ is determined by post-selecting data from the second measurement $M_1$ in which the qubit was in state $|0\rangle$ in the first measurement $M_0$. In addition, we measured the thermal population of each qubit biased at the idle point by a DC current. The measured thermal populations $P_{\mathrm{th}}^{\mathrm{idle}}$ at the idle

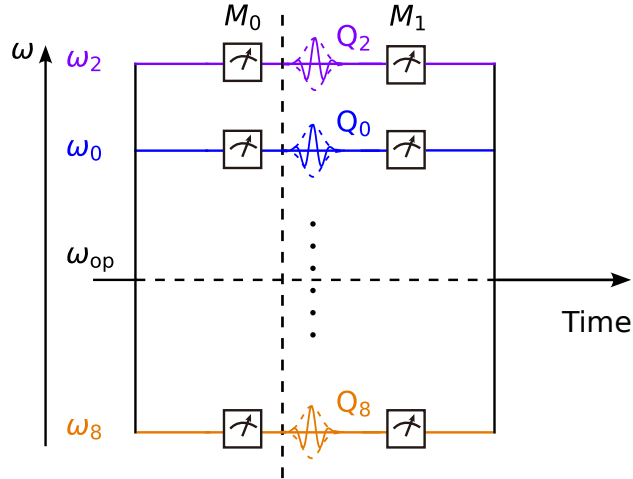

FIG. S7. **Pulse sequence for evaluating assignment fidelities:** All the qubits are $Z$-biased at the operation frequency $\omega_{\mathrm{op}}$ for initialization for a long period (600 $\mu$s) to allow the system to reach a steady state. Then, we apply the $Z$-pulses to shift all the qubits to their idle frequencies. Subsequently, we perform the first measurement $M_0$, apply qubit drives, and conduct the second measurement $M_1$, sequentially.

points are summarized in Table I. From these thermal-population values, the effective qubit temperature was calculated [15] and shown in Fig. S6c. More parameters of the chip can be found in Table I and Figs. S5 and S6.

## Appendix E: System calibration

In the experiments, all the qubits are initially biased at the operation frequency with the $Z$-control DC bias ($Z$-bias), while the initial-state preparation and the final-state readout are performed at the idle points by applying $Z$-pulses. To achieve optimal energy-relaxation times $T_1$ across all qubits, minimize the impact of defects and $ZZ$ coupling, and achieve high-fidelity multi-qubit initialization and readout, the following parameters need to be carefully optimized or calibrated: (i) $Z$-pulse distortion, (ii) qubit frequency setup, (iii) single-qubit gates, (iv) timing alignment between different control channels, (v) qubit frequency alignment, (vi) the phase of qubit, and (vii) qubit frequency over time . The qubit-frequency drift is evaluated with the Ramsey decay measurement and then corrected with the compensating $Z$-bias. For timing alignment and single-qubit gate optimization, we follow the methods in Ref. [11]. Further details regarding the optimization and calibration of other parameters are provided below.

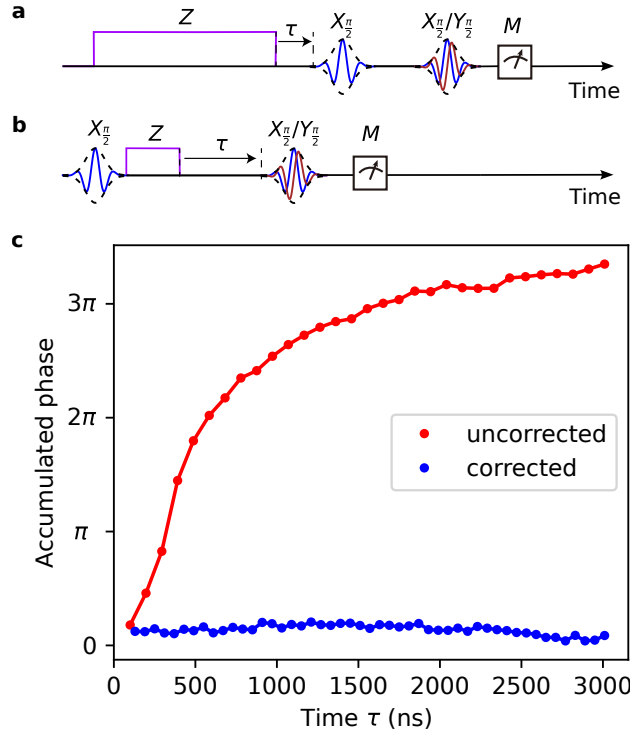

FIG. S8. **Z-pulse distortion calibration**. **a**, Pulse sequence to calibrate the long-time-scale Z-pulse distortion. **b**, Pulse sequence to calibrate the short-time-scale Z-pulse distortion. **c**, Accumulated phase measured by the pulse sequence in **a** before and after the Z-pulse distortion calibration.

### 1. *Z-pulse distortion calibration*

In our experiments, the Z-pulse distortion would significantly affect the fidelities of gates and readout since they are performed at the idle points with Z-pulses applied. Any component with limited bandwidth, impedance mismatch, or parasitic capacitance and inductance in the control line can contribute to the Z-pulse distortion. As shown in Fig. S3a, we use spring contacts to deliver the control signals to the ports on the chip. By employing the impedance-matched design, we can effectively suppress the distortion of the Z-pulse on the package side.

To calibrate the Z-pulse distortion, we need to bias each qubit to a point where its frequency is highly sensitive to the Z-pulse amplitude. In our case, the qubit is biased below 5.8 GHz, with the Z-pulse amplitude significantly larger than that used in the experiments. Since the tunable transmon qubit exhibits shorter $T_2$ at the frequency-sensitive point, the standard Ramsay pulse sequence to measure the phase shift due to the Z-pulse Fig. S8b [16] cannot be used for long-time-scale distortion due to the large dephasing rate. Therefore, we first calibrate the distortion of the Z-pulse using the waveform sequence shown in Fig. S8a [11] by applying a long-time (10 $\mu$s) Z-pulse. After a delay time $\tau$, we apply two half-$\pi$ pulses with

a constant time spacing (200 ns) to measure the accumulated phase. The longtime $Z$-pulse can effectively separate the distortion due to the rising edge from that due to the falling edge, simplifying the calibration process. Then, we further characterize the short-time-scale $Z$-pulse distortions with the waveform sequence in Fig. S8b, as the accumulated phase measured in this waveform is more sensitive to the initial portion of the distorted $Z$-pulse. Finally, we find that applying several exponential-decay corrections to the ideal $Z$-pulse is sufficient to correct the distortion of the $Z$-pulse for all the qubits. In Fig. S8c, we present the accumulated phase of qubit $Q_5$ measured by using the sequence in Fig. S8a with corrected and uncorrected $Z$-pulses, respectively.

### 2. Frequency setup

Several considerations must be taken into account when optimizing the frequency setup, as outlined below:

1. **Energy relaxation**: Due to unwanted coupling with two-level systems (TLSs) or other modes, the transmon qubits typically exhibit $T_1$ fluctuations as a function of the qubit frequency and time [17, 18]. Consequently, we measure the $T_1$ distribution for each qubit in the frequency range between the operation frequency and the idle frequency, ensuring that none of the qubits interact with the loss modes.

2. **Qubit–qubit coupling at the idle points**: In our chip, the coupling strengths between nearest-neighbor (NN) qubits range from 2.9 to 4.8 MHz. After optimizing the idle frequencies of all the qubits, the residual $XY$ coupling between all the NN qubits can be suppressed below 0.06 MHz. Additionally, we ensure that the minimum frequency detuning between the next-nearest-neighbor (NNN) qubits is larger than 5 MHz to suppress any unwanted qubit drive from the $X$-crosstalk, which should be below −30 dB in our system. Finally, to avoid the potential two-photon excitations due to the $X$-crosstalk, we also avoid exact frequency matches with $\omega_{02}/2$ between the neighboring qubits, where $\omega_{02}$ is the frequency difference between the ground state and the second excited state.

3. **$ZZ$ coupling**: For coupled two neighboring transmon qubits, because of the weak anharmonicities, there is non-negligible $ZZ$ coupling due to interaction through higher-energy levels such as $|11\rangle$, $|02\rangle$, and $|20\rangle$ [7]. This coupling can be expressed as

$$\Omega_{ZZ} = -\frac{2g^2(\eta_1 + \eta_2)}{(\Delta - \eta_1)(\Delta + \eta_2)}, \tag{S1}$$

where $\eta_1$ and $\eta_2$ represent the anharmonicity of the qubits, $g$ is the their coupling strength, and $\Delta$ denotes the difference in their frequencies. The $ZZ$ interaction introduces inter-qubit control crosstalk during parallel quantum operations, thereby reducing the fidelity of multi-qubit state initialization. Furthermore, this parasitic coupling degrades the fidelity of simultaneous readout for the multi-qubit state. Consequently, the final state of the quantum system continues to evolve under the non-negligible $ZZ$ coupling. By optimizing the frequency setup, as shown in Fig. S6g, we suppressed the $ZZ$ coupling between any NN qubits to below 110 KHz.

4. **Dephasing**: In our chip, $T_1$ is always much larger (>10 μs) than the system evolution time for the experiment (300 ns). However, the short dephasing time (as shown in Fig. S5) could significantly influence the experiment. The qubits may couple to some modes with thermal population (TLSs, etc.), resulting in a significantly larger dephasing rate, which can lead to reduced purity in the multi-qubit states. By measuring $T_2^*$ of each qubit within a frequency range of several tens of MHz, both at the operation point and the idle points, we might find the undesired coupling modes. To avoid these modes, we adjust the operation point or the idle points of the affected qubits, ensuring that they are decoupled from these modes. After this optimization, all the qubits work well at the operation point and their idle points.

### 3. Qubit-frequency alignment

In Ref. [11], the frequency alignment was optimized by the vacuum Rabi oscillations between the nearest-neighbor qubits. Then, the $Z$-pulses, corrected by the crosstalk matrix, were applied to shift the qubits from the idle points to the operation point. However, in this experiment, we employed an alternative approach to align the frequencies of all the qubits, omitting the $Z$-pulse correction by the crosstalk matrix, as the current chip exhibits $Z$-crosstalk nearly ten times smaller. The details of the calibration process are as follows:

1. Bias each qubit to the operation point using $Z$-bias, while detuning the frequencies of other qubits by more than 400 MHz. Next, perform the Ramsey measurement to fine-tune the $Z$-bias voltage $V_{DC}$, adjusting each qubit to the operation frequency.

2. Use the $Z$-pulse to shift each qubit from the operation point calibrated in step 1 to its idle point. Then, perform the Ramsey measurement at the idle point. Fine-tune the $Z$-pulse amplitude $V_{pulse}$, setting each qubit to its target idle frequency.

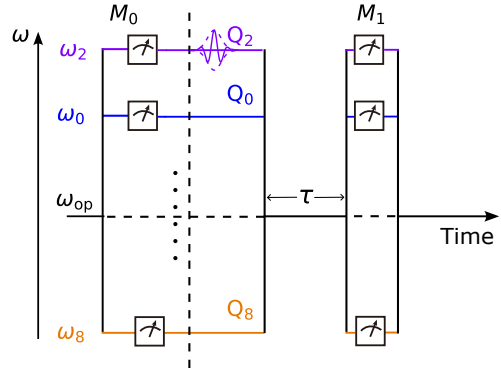

FIG. S9. **Pulse sequence to probe deep thermalization**. All the qubits are initially aligned to the operation point. Then, we turn on the $Z$-pulses, perform the readout $M_0$ and prepare the initial state. Subsequently, we turn off the $Z$-pulses, allowing the system to evolve at the operation point for a duration of $\tau$. Finally, the $Z$-pulses for all the qubits are turned on to freeze the system evolution and perform the readout $M_1$.

3. Bias all the qubits at the operation point by $Z$-bias with the calibrated $Z$-bias voltage $V_{\mathrm{DC}}$ in step 1. Next, perform the Ramsey measurement for each qubit at its idle point with the calibrated $Z$-pulse amplitude $V_{\mathrm{pulse}}$ in step 2. If any qubits experience frequency shifts due to the $Z$-bias crosstalk, then, fine-tune the $Z$-bias voltage for those qubits to compensate for the frequency shifts, successfully calibrating the $Z$-bias crosstalk.

4. After calibrating each qubit's $Z$-bias voltage $V_{\mathrm{DC}}$ in step 3, we perform the Ramsey measurement on each qubit with all the qubits' $Z$-pulses applied. Due to $Z$-pulse crosstalk, some qubits may exhibit frequency shifts at their idle points. We then fine-tune the $Z$-pulse amplitudes of these qubits to compensate for the frequency shifts, successfully calibrating the $Z$-pulse crosstalk.

In the process of the $Z$-crosstalk calibration, we found that most qubits experienced frequency shifts of less than 1 MHz from the crosstalk of the $Z$-bias and the $Z$-pulse, signifying minimal $Z$-crosstalk within our system. As all calibrations above rely on the Ramsey-oscillation measurements, the frequency alignment precision is limited by the fitting error of the Ramsey oscillation, which is generally below 10 kHz. After calibration, we can mitigate the effects of the $Z$-crosstalk and ensure that each qubit functions at its target frequency during the experiments. As shown in Fig. S10, the experimental results for the 12Q chain system are in good agreement with the theoretical simulations.

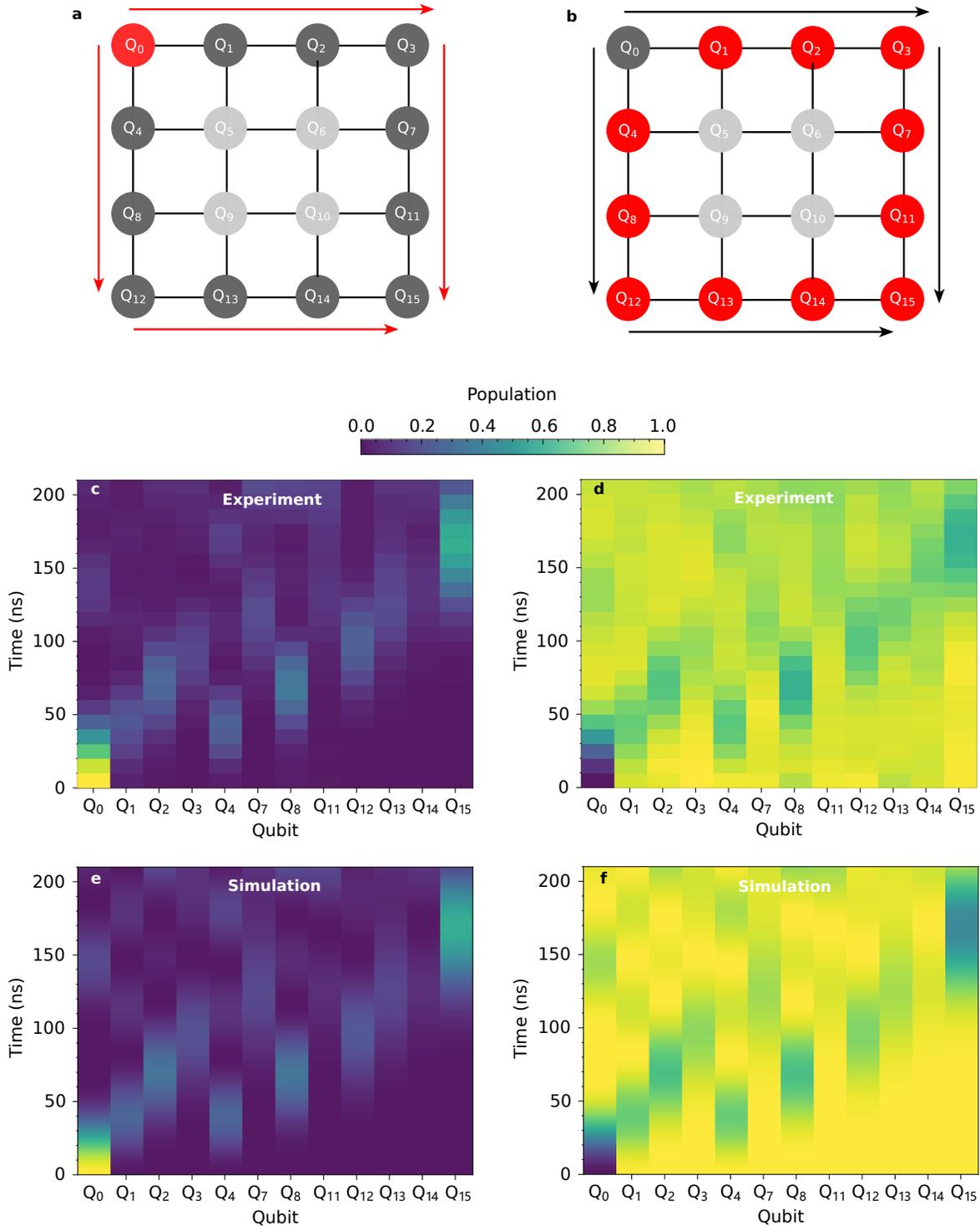

FIG. S10. **Dynamics of particle and hole propagating along a 12-qubit loop**. **a**, $Q_0$ is initially excited to the state $|1\rangle$ while all other qubits are in state $|0\rangle$, creating a localized particle which propagates along the 12-qubit loop. **b**, All the qubits except $Q_0$ are initially excited to the state $|1\rangle$, leaving $Q_0$ in $|0\rangle$ to form a 'hole'. **c** and **d**, Experimental results showing particle (c) and hole (d) propagations under the initial conditions of **a** and **b**, respectively. **e** and **f**, Simulations of the particle and hole propagations, respectively. The agreement between the experimental results and the simulations demonstrates that the system can be accurately modeled by using an XY spin model. Furthermore, it confirms the precise synchronization of the operations across the qubits.

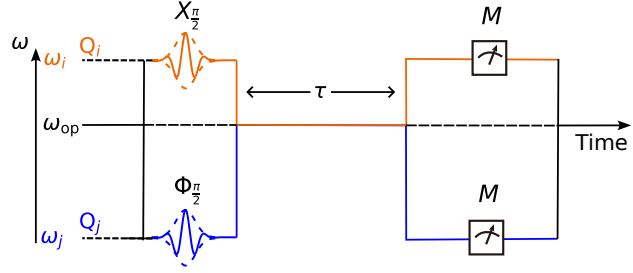

FIG. S11. **Pulse sequence for calibrating the relative phase**. We rotate $Q_i$ and $Q_j$ to the $XY$ plane and set the initial phase of $Q_i$ as the reference. Then, we adjust the phase of $Q_j$, turn off the $Z$-pulse to let them resonantly couple for a time duration $\tau = \pi/4g_{ij}$, where $g_{ij}$ is the coupling strength between $Q_i$ and $Q_j$. At the end, we turn on the $Z$-pulse and perform the readout.

### 4. Relative-phase calibration

In the information leakage benchmark experiment, as illustrated in Fig. 4a of the main text, all qubits in the $3 \times 3$ array are initially prepared in a superposition state. Specifically, the center and corner qubits are prepared in the state $|X_+\rangle = \frac{1}{\sqrt{2}}(|0\rangle + |1\rangle)$, while the remaining qubits are prepared in the state $|Y_+\rangle = \frac{1}{\sqrt{2}}(|0\rangle + i|1\rangle)$. To prepare this multi-qubit state, we must calibrate the relative phases between all the qubits [19]. The Hamiltonian for the coupled two-qubit system can be written as

$$\hat{H}/\hbar = \frac{1}{2}\omega\hat{\sigma}_z^i + \frac{1}{2}\omega\hat{\sigma}_z^j + g\left(\hat{\sigma}_+^i\hat{\sigma}_-^j + \hat{\sigma}_-^i\hat{\sigma}_+^j\right). \tag{S2}$$

Here, we assume that the two qubits have the same frequency $\omega$, and $g$ is the coupling strength between them. As shown in Fig. S11, in the relative-phase calibration of the two qubits, we use the initial phase of $Q_i$ as a reference, preparing it to the state $|\psi_i\rangle = \frac{1}{\sqrt{2}}(|0\rangle + |1\rangle)$, then we prepare another qubit to the state $|\psi_j\rangle = (|0\rangle + e^{i\phi}|1\rangle)$ with relative phase $\phi$. The initial state for the two-qubit system can be written as $|\psi_{ij}(t=0)\rangle = \frac{1}{2}(|0\rangle + |1\rangle) \otimes (|0\rangle + e^{i\phi}|1\rangle)$. Under Eq. (S2), we obtain the state evolution over time as follows:

$$\begin{aligned}
|\psi_{ij}(t)\rangle =& \frac{1}{2}e^{-i\omega t}|00\rangle + \frac{1}{2\sqrt{2}}e^{i\phi}(|\Psi_+(t)\rangle - |\Psi_-(t)\rangle) + \frac{1}{2\sqrt{2}}(|\Psi_+(t)\rangle + |\Psi_-(t)\rangle) + \frac{1}{2}e^{i\omega t}|11\rangle \\
=& \frac{1}{2}e^{-i\omega t}|00\rangle + \frac{1}{4}\left[(e^{-igt} - e^{igt}) + e^{i\phi}(e^{-igt} + e^{igt})\right]|01\rangle + \frac{1}{4}\left[(e^{-igt} + e^{igt}) + e^{i\phi}(e^{-igt} - e^{igt})\right]|10\rangle \\
&+ \frac{1}{2}e^{i\omega t}|11\rangle.
\end{aligned} \tag{S3}$$

Here, $|\Psi_+\rangle = \frac{1}{\sqrt{2}}(|01\rangle + |10\rangle)$ and $|\Psi_-\rangle = \frac{1}{\sqrt{2}}(|01\rangle - |10\rangle)$ are the eigenstates of the coupled qubits in

Eq. (S2). Then, the probabilities for observing the four energy eigenstates $|00\rangle$, $|11\rangle$, $|01\rangle$ and $|10\rangle$ evolve as follows:

$$P_{00}(t) = \frac{1}{4}, \tag{S4}$$

$$P_{11}(t) = \frac{1}{4}, \tag{S5}$$

$$P_{01}(t) = \left| \frac{1}{4} \left[ \left( e^{-igt} - e^{igt} \right) + e^{i\phi} \left( e^{-igt} + e^{igt} \right) \right] \right|$$
$$= \frac{1}{4} \left[ 1 - \sin(\phi)\sin(2gt) \right], \tag{S6}$$

$$P_{10}(t) = \left| \frac{1}{4} \left[ \left( e^{-igt} + e^{igt} \right) + e^{i\phi} \left( e^{-igt} - e^{igt} \right) \right] \right|$$
$$= \frac{1}{4} \left[ 1 + \sin(\phi)\sin(2gt) \right]. \tag{S7}$$

From Eqs. (S4)–(S7), we expect that the probabilities for states $|00\rangle$ and $|11\rangle$ are independent of the phase and time, each remaining constant at $1/4$. This is consistent with the experimental results presented in Fig. S12b. When the phase is neither 0 nor $\pi$, the probabilities $P_{01}$ and $P_{10}$ will oscillate with the phase $\phi$ of $Q_j$, ranging from $\frac{1}{4}[1 - \sin(2gt)]$ to $\frac{1}{4}[1 + \sin(2gt)]$. In the relative-phase calibration, we fixed the evolution time to be $\pi/4g$ for better SNR. The position of the black arrow in Fig. S12b marks the phase offset between $Q_3$ and $Q_7$, where $P_{01} = 0$ with a negative slope. We calibrated the necessary qubit pairs to determine the relative-phase offsets between all the qubits in the $3 \times 3$ array. The results are presented in Fig. S12a.

### 5. Dynamical-phase calibration

In the information-leakage benchmark experiment, we measured the expectation value $\langle X \rangle$ of all the qubits (Fig. 4a) with a $X_{\frac{\pi}{2}}$ pulse applied to each qubit before the measurement. As the system evolves with all the qubits at the operation point while being characterized at their idle points, all the qubits will accumulate a phase during the system evolution. We use the sequence in Fig. S13 to calibrate each qubit's dynamical phase. The qubit is prepared in the state $\frac{1}{\sqrt{2}}(|0\rangle + |1\rangle)$ at the idle point with the $Z$-pulse applied. Then, the $Z$-pulse is turned off, tuning the qubit to the operation point. After a time $\tau$, the qubit is returned to the idle point, followed by an $X$ rotation, and finally measured. The probability of measuring the qubit at state $|1\rangle$ oscillates when we sweep the phase of the final $X$ rotation, allowing us to determine the accumulated dynamical phase at which the qubit reaches maximum probability at state $|1\rangle$. Specifically, the qubit accumulates its phase in three stages during the system evolution: the falling edge ($0 < t < \tau_{\mathrm{f}}$), the flat section ($\tau_{\mathrm{f}} < t < \tau_{\mathrm{r}}$) and the rising edge ($\tau_{\mathrm{f}} < t < \tau$).

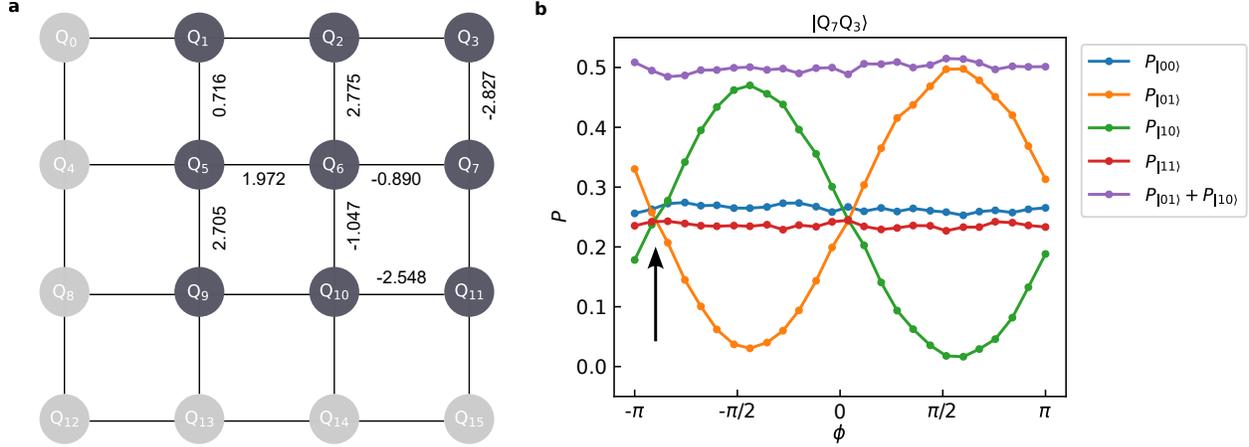

FIG. S12. **Relative-phase calibration**. **a**, Relative-phase offsets (rad) between the adjacent qubits. **b**, Results of the experiments in Fig. S11. The position of the black arrow indicates the relative-phase offset between Q$_3$ and Q$_7$

$$\phi(\tau) = \int_0^{\tau_{\mathrm{f}}} \Delta\omega_{\mathrm{r}}(t)\,dt + \int_{\tau_{\mathrm{f}}}^{\tau_{\mathrm{r}}} \Delta\omega_{\mathrm{r}}(t)\,dt + \int_{\tau_{\mathrm{r}}}^{\tau} \Delta\omega_{\mathrm{r}}(t)\,dt. \tag{S8}$$

Finally, we sweep the evolution time $\tau$, performing the $X_{\frac{\pi}{2}}$ rotation to the final state with the calibrated phase. As shown in Fig. S13b, the phases of all qubits' final states for each evolution time $\tau$ are well calibrated, and the decay rate of the qubit expectation values $\langle X \rangle$ agrees well with the qubit dephasing time $T_2^*$ illustrated in Fig. S5h.

## 6. Long-time readout stability

In our chip, four readout resonators in a single unit are coupled to a single band-pass filter. By applying frequency-multiplexed heterodyne-mixing signals to the filter, we realize multiplexed qubit readouts through the four resonators. In the readout optimization, we fine-tune the readout frequencies, durations, and amplitudes to maximize the separation between the readout signals conditioned on the qubits' ground and excited states in the IQ plane. The hyperplane parameters are calculated to maximize the probability of correctly classifying the states $|0\rangle$ and $|1\rangle$. The readout quality is quantified by the readout fidelity, defined as

$$F_{\mathrm{RO}} = \frac{1}{2}(F_{00} + F_{11}), \tag{S9}$$

where $F_{00}$ ($F_{11}$) is the probability of correctly classifying the qubit state $|0\rangle$ ($|1\rangle$) as 0 (1).

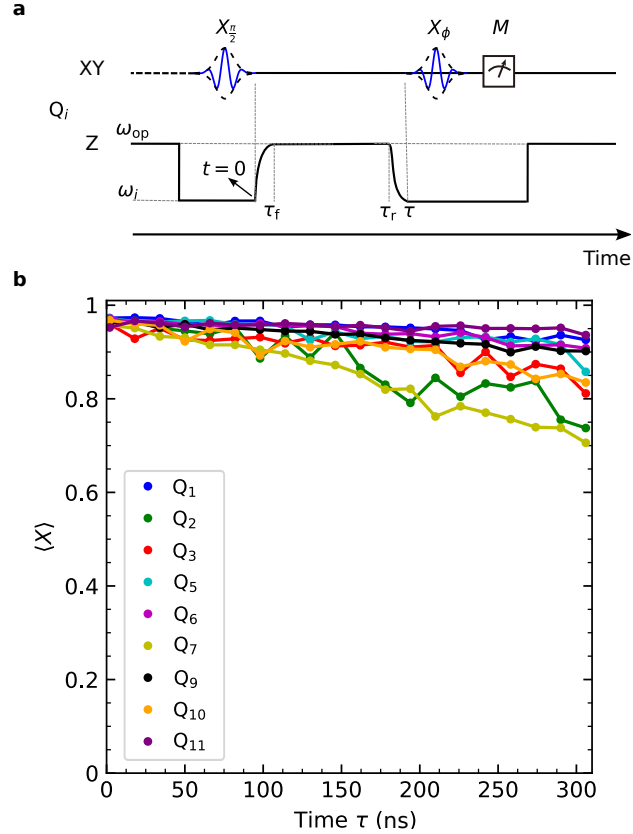

FIG. S13. **Dynamical-phase calibration**. **a**, Pulse sequence to calibrate the accumulated dynamical phase during the $Z$-pulse control. **b**, Expectation values $\langle X \rangle$ after calibration for each qubit as a function of time $\tau$.

The readout pulse length for each readout resonator is approximately 400 ns, with a demodulation window duration ranging from 520 ns to 600 ns, primarily determined by the external coupling rate of each resonator, $\kappa$ (as illustrated in Fig. S5h). Figure S14a illustrates a set of the single-shot readout data for $Q_{12}$. The data is acquired by preparing all qubits in the states $|0\rangle^{\otimes 16}$ (shown in blue) and $|1\rangle^{\otimes 16}$ (shown in red). The dashed line is the principal axis used to distinguish the qubit states $|0\rangle$ and $|1\rangle$. Figure S14b shows histograms of the data in Fig. S14a projected along the principal axis in the IQ plane. The dot-dashed lines represent the Gaussian fits to the signal distributions of states $|0\rangle$ and $|1\rangle$. For this dataset, the readout assignment fidelity is $F_{\rm RO} = 99.68\%$, with $F_{00} = 99.87\%$ and $F_{11} = 99.49\%$. The difference between $F_{11}$ and $F_{00}$ is attributed to the energy relaxation of the qubit during the readout.

In our experiments, we take the data continuously for a long time. To maintain the system stability, for every 3.5 hours, we calibrated each qubit's frequency and $\pi$-pulse amplitude and then characterized the readout fidelity without modifying any readout parameters. In Fig. S15, we present the results of readout data collected over five days of measurement. For most of the qubits, the readout remained highly stable, with variations in $F_{ii}$ ($i \in \{0, 1\}$) below 0.02. Over five days of continuous measurement, we achieved an

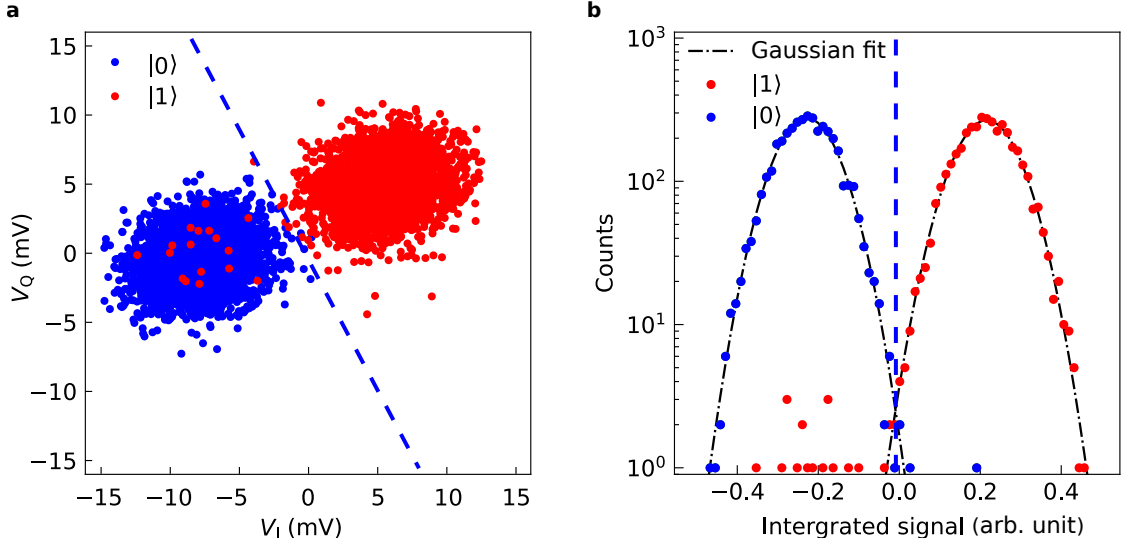

FIG. S14. **Single-shot readout. a**, Time-integrated IQ signals of $Q_{12}$ when all the sixteen qubits are prepared in state $|0\rangle^{\otimes 16}$ and in state $|1\rangle^{\otimes 16}$. **b**, Histograms of the data from **a**, showing data projected along the principal axis in the IQ plane. Dot-dashed lines represent Gaussian fits to the signal distributions for $|0\rangle$ and $|1\rangle$.

average readout fidelity $F_{RO}$ of 98.51% across all the qubits, with average fidelities of 99.57% for $F_{00}$ and 97.46% for $F_{11}$.

## Appendix F: Benchmark of many-body information leakage

In this section, we explore the universality of our method for benchmarking many-body information leakage. When the system is coupled to the environment, it transitions into a mixed state. The corresponding density matrix can be expressed as a sum of orthogonal pure states:

$$\hat{\rho} = \sum_j \lambda_j |\phi_j\rangle \langle\phi_j|, \tag{S1}$$

where $\langle\phi_i|\phi_j\rangle = \delta_{ij}$ and $\sum_j \lambda_j = 1$. Without loss of generality, we assume $\lambda_1 \geq \lambda_2 \geq \lambda_3 \ldots$. A large degree of information leakage results in low purity $p$ of $\hat{\rho}$, defined as

$$p := \text{Tr}\hat{\rho}^2 = \sum_j \lambda_j^2, \tag{S2}$$

and is related to the second-order Rényi entropy:

$$S_2 = -\ln \text{Tr}\hat{\rho}^2 = -\ln p. \tag{S3}$$

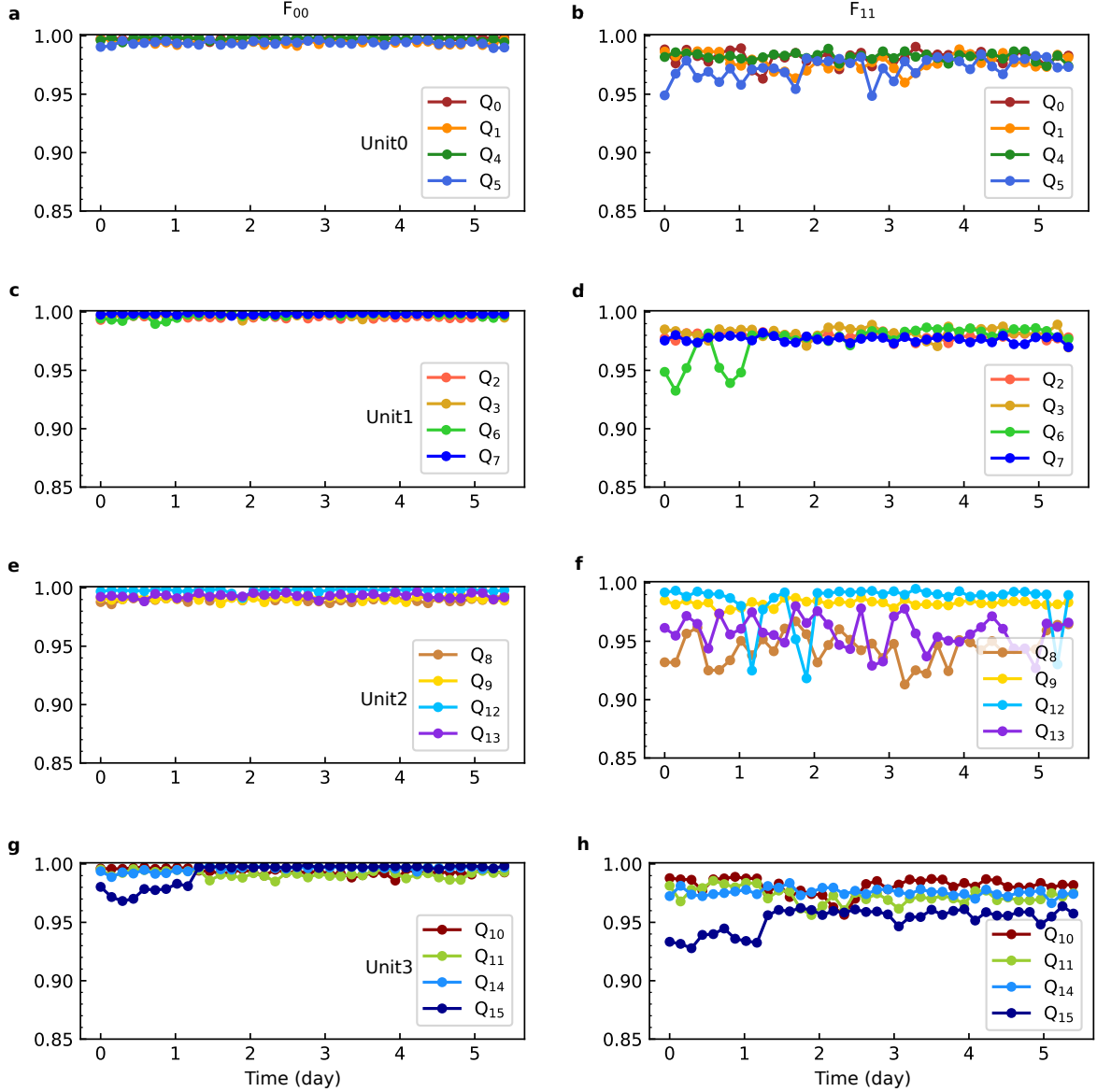

FIG. S15. **Simultaneous readout fidelities $F_{00}$ and $F_{11}$, of the 16 qubits in over five-day continuous measurement.**

We consider the case of small information leakage, where $\lambda_1 \gg \sum_{j=2} \lambda_j$. Under this condition, the second-order Rényi entropy can be approximated as

$$S_2 = -\ln\left(\sum_j \lambda_j^2\right) \sim -2\ln\lambda_1. \tag{S4}$$

Now we focus on the projected ensemble of the density matrix $\hat{\rho}$. Each projected state of subsystem A can be expressed as

$$\hat{\rho}_A(z_B) = \sum_j \lambda_j |\phi_{A,j}(z_B)\rangle \langle\phi_{A,j}(z_B)|, \tag{S5}$$

where $|\phi_{\mathrm{A},j}(z_{\mathrm{B}})\rangle$ represents the projected state corresponding to $|\phi_j\rangle$:

$$|\phi_{\mathrm{A},j}(z_{\mathrm{B}})\rangle = \mathcal{A}(\mathbb{I}_{\mathrm{A}} \otimes \langle z_{\mathrm{B}}|)|\phi_j\rangle,\tag{S6}$$

with $\mathcal{A}$ as the normalization factor. In a quantum many-body system, after sufficient time evolution, $|\phi_{\mathrm{A},j}(z_{\mathrm{B}})\rangle$ can be treated as a random state. Using the Bloch-sphere representation, $\hat{\rho}_{\mathrm{A}}(z_{\mathrm{B}})$ can be regarded as the sum of a series of random vectors. In the case of small information leakage, the length of these vectors (equivalent to the purity) is proportional to $\lambda_1$. Consequently, the ensemble-averaged entropy of $\hat{\rho}$ can be approximated as

$$\bar{E}_{\mathrm{A}} = -\sum_{z_{\mathrm{B}}} p(z_{\mathrm{B}}) \ln \mathrm{Tr}\hat{\rho}_{\mathrm{A}}^2(z_{\mathrm{B}}) \sim -\ln \lambda_1.\tag{S7}$$

From Eqs. (S4) and (S7), we can find that $\bar{E}_{\mathrm{A}}$ is proportional to the second-order Rényi entropy $S_2$. This demonstrates that $\bar{E}_{\mathrm{A}}$ **serves as an effective measure of information leakage in quantum many-body systems**.

To validate the above discussions, we perform numerical simulations for both Markovian noise and $1/f$ noise. We consider three representative models: the 1D mixed Ising model, the effective Hamiltonian for Rydberg atoms with long-range van der Waals interaction, and the XY model. The corresponding Hamiltonians are written as:

$$\hat{H}_{\mathrm{MIsing}} = \sum_{j}^{L-1} \hat{\sigma}_j^z \hat{\sigma}_{j+1}^z + h_z \sum_{j=1}^{L} \hat{\sigma}_j^z + h_x \sum_{j=1}^{L} \left(\hat{\sigma}_j^x - \hat{\sigma}_1^z - \hat{\sigma}_L^z\right),\tag{S8}$$

$$\hat{H}_{\mathrm{RA}} = V \sum_{i>j}^{L-1} \frac{\hat{n}_j \hat{n}_{j+1}}{|i-j|^6} - \Delta \sum_{j=1}^{L} \hat{n}_j + \Omega \sum_{j=1}^{L} \hat{\sigma}_j^x,\tag{S9}$$

$$\hat{H}_{\mathrm{XY}} = \sum_{\langle i,j \rangle} J_{ij}(\hat{\sigma}_i^+ \hat{\sigma}_j^- + \mathrm{h.c.}).\tag{S10}$$

In addition, without loss of generality, we only consider the infinite-temperature systems.

### 1. Markovian noise

We first consider Markovian noise, where the system dynamics are governed by the Lindblad master equation:

$$\frac{d}{dt}\hat{\rho} = -i[\hat{H}, \hat{\rho}] + \sum_{j} \gamma_j \left(\hat{L}_j \hat{\rho} \hat{L}_j^\dagger - \frac{1}{2}\{\hat{L}_j^\dagger \hat{L}_j, \hat{\rho}\}\right).\tag{S11}$$

Here, $\gamma_j$ represents the dissipation rate (the coupling strength between the system and the environment), and $\hat{L}_j$ is the jump operator.

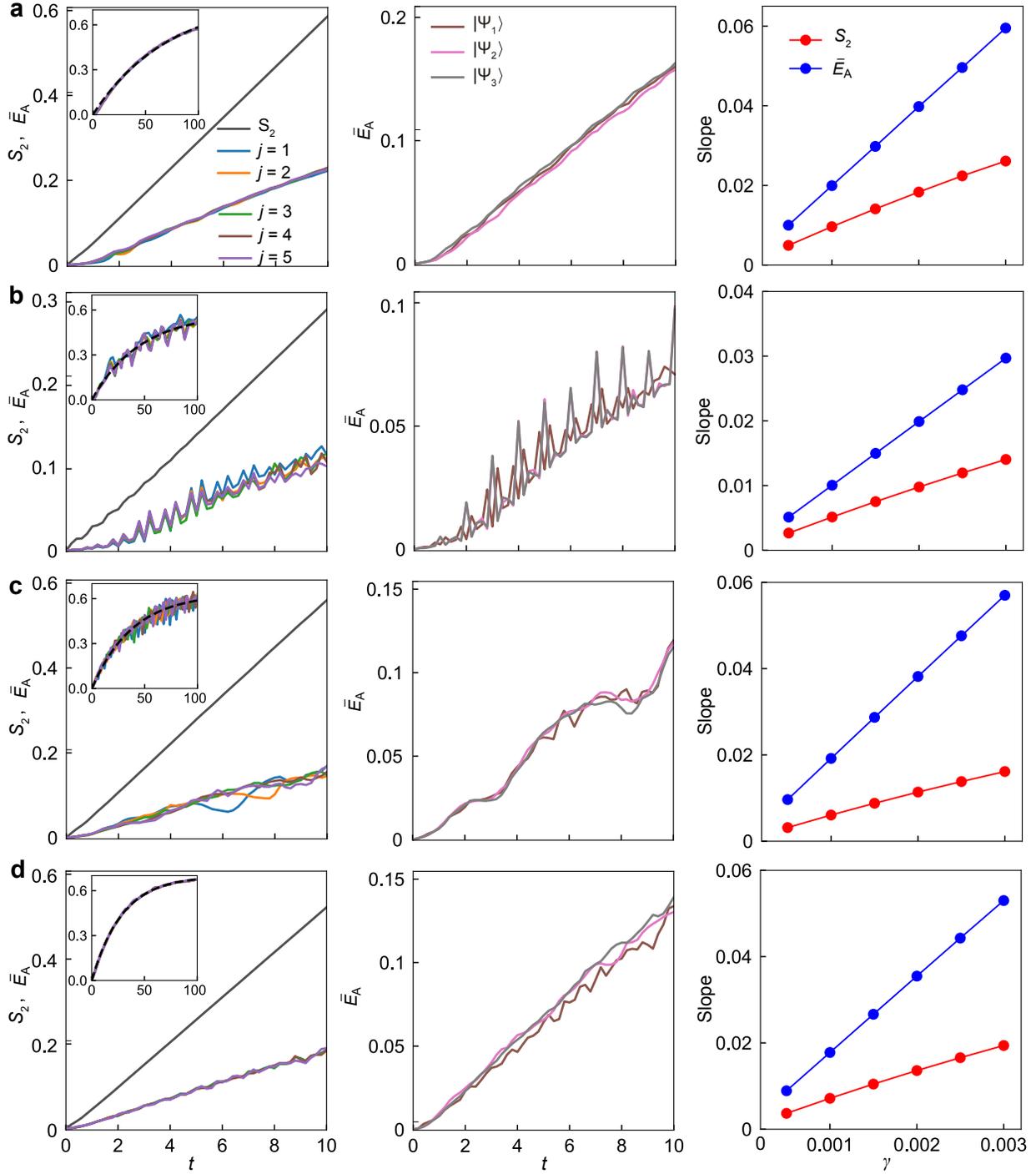

FIG. S16. **Benchmark of many-body information leakage. a,** 1D mixed Ising model. **b,** Interacting Rydberg-atom Hamiltonian. **c,** 1D XY model. **d,** 2D XY model. (Left panel) Dynamics of $S_2$ and $\bar{E}_A$, where we choose site $j$ as subsystem A. The initial states are corresponding to $|\Psi_1\rangle$, and $\gamma = 0.002$. The insets show the long-time evolution of $\bar{E}_A$, and the black dashed curves are fits according to Eq. (S12). (Middle panel) Dynamics of $\bar{E}_A$ for different initial states with site 5 as subsystem A and $\gamma = 0.002$. (Right panel) Slopes of $S_2(t)$ and $\bar{E}_A(t)$ in **a** versus the dissipation rate $\gamma$.

For 1D mixed Ising model, we choose the parameters $h_z = 0.8090$, $h_x = 0.9045$, and $L = 10$, and also consider the dephasing channel $\hat{L}_j = \hat{\sigma}_j^z$. In addition, we consider three initial states: $|\Psi_1\rangle = |Y_+Y_+Y_+...\rangle$, $|\Psi_2\rangle = |Y_+Y_-Y_+Y_-...\rangle$, and $|\Psi_3\rangle = |X_+X_-X_+X_-...\rangle$.

For $\hat{H}_{\text{RA}}$, we choose the parameters $V = 1.44$, $\Delta = 0.9$, $\Omega = 4.7$, and $L = 10$, and consider the energy-relaxation channel $\hat{L}_j = \hat{\sigma}_j^-$. We consider three initial states: $|\Psi_1\rangle = |Y_+Y_+Y_+Y_+...\rangle$, $|\Psi_2\rangle = |0000...\rangle$, and $|\Psi_3\rangle = |0101...\rangle$.

For the XY model, we consider both 1D and 2D systems with the dephasing channel $\hat{L}_j = \hat{\sigma}_j^z$, and the system sizes are $L = 10$ and $3 \times 3$, respectively. In the 1D case, we consider three initial states: $|\Psi_1\rangle = |X_+Y_+X_+Y_+...\rangle$, $|\Psi_2\rangle = |X_+X_+X_-X_-X_+X_-X_-X_+X_+\rangle$, and $|\Psi_3\rangle = |X_+X_+Y_+Y_-X_+X_+Y_+Y_-X_+X_+\rangle$. In the 2D case, we have $|\Psi_1\rangle = |X_+Y_+X_+Y_+...\rangle$, $|\Psi_2\rangle = |X_+X_+X_+X_-X_-X_-X_+X_+X_+\rangle$, and $|\Psi_3\rangle = |X_+X_-X_-Y_+Y_-Y_-X_+X_-X_-\rangle$.

The numerical results are presented in Fig. S16. The left panels demonstrate that $\bar{E}_\text{A}$ exhibits similar dynamics across different subsystems A. In the short-time regime ($t \ll 1/\gamma$), both $S_2$ and $\bar{E}_\text{A}$ show a linear increase. Over a longer time scale, the dynamics of $\bar{E}_\text{A}$ can be well approximated by

$$\bar{E}_\text{A} = E_0\left(1 - e^{-t/\tau_\text{MB}}\right), \tag{S12}$$

where $E_0$ and $\tau_\text{MB}$ characterize the steady-state value and relaxation time, respectively. The middle panels reveal that $\bar{E}_\text{A}$ is largely independent of the initial states. We also obtain the slopes of $S_2(t)$ and $\bar{E}_\text{A}(t)$ in the short-time regime by using a linear fit. The slopes versus the dissipation rate $\gamma$ are plotted in the right panels of Fig. S16. The results indicate that the slopes are proportional to $\gamma$.

## 2. $1/f$ noise

In many quantum simulators, dephasing mainly results from $1/f$ noise (e.g., superconducting qubits, nuclear spins, Rydberg atoms etc.), which is not Markovian. Now we focus on $1/f$ noise, where the system dynamics are governed by the Hamiltonian:

$$\hat{H}(t) = \hat{H}_0 + \frac{1}{2}\sum_j \xi_j(t)\hat{\sigma}_j^z, \tag{S13}$$

$$\langle \xi_j(0)\xi_k(t)\rangle = \int_{-\infty}^{\infty} \frac{d\omega}{2\pi} S_{jk}(\omega)e^{-i\omega t}. \tag{S14}$$

where $\xi(t)$ is a Gaussian fluctuation, and $S_{jk}(\omega)$ is the spectral density of the noise. Here, we do not consider the spatial correlation of the noise, so the noise spectral density of noise satisfies

$$S_{jk}(\omega) = \frac{\delta_{jk}A_j}{|\omega|}, \tag{S15}$$

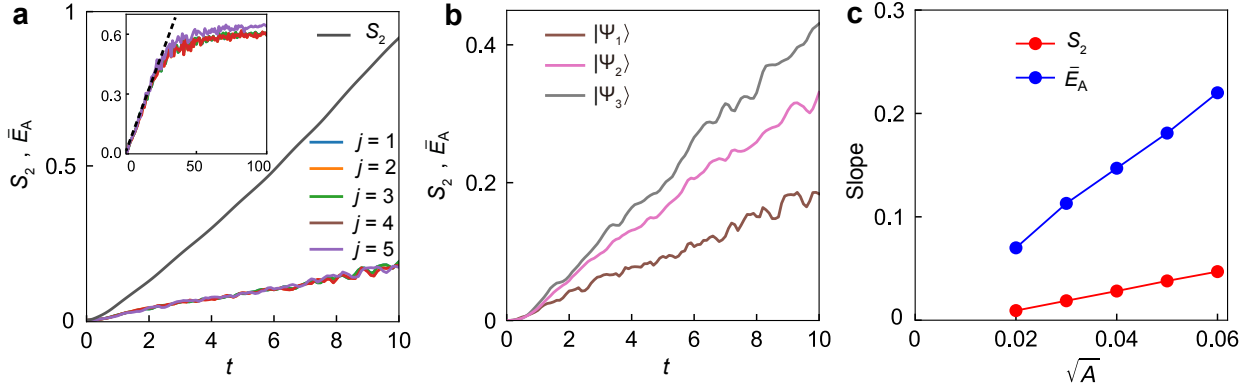

FIG. S17. **Benchmark of many-body information leakage due to $1/f$ noise in the 2D XY model. a,** Dynamics of $S_2$ and $\bar{E}_A$ under $1/f$ noise. We choose site $j$ as subsystem A. The initial states are corresponding to $|\Psi_1\rangle$, and $\sqrt{A_j} = 0.03$. The insets are the long-time evolution of $\bar{E}_A$, and the black dashed curve is a linear fit. **b,** Dynamics of $\bar{E}_A$ for different initial states with site 5 as subsystem A and $\sqrt{A_j} = 0.03$. **c,** Slopes of $S_2(t)$ and $\bar{E}_A(t)$ in **a** versus the noise intensity $\sqrt{A_j}$.

where $A_j$ is the noise strength. Generally, the low-frequency regime of $1/f$ noise dominates the dephasing in the system, i.e., $\omega_h \ll 1/T$, where $\omega_h$ is the high-frequency cutoff and $T$ is the total evolution time. Thus, Eq. (S15) can be considered as a time-independent system with disorder.

For the numerical simulation, we consider a 2D XY model, i.e., $\hat{H}_0 = \hat{H}_{XY}$. We choose the low- and high-frequency cutoff as $\omega_l = 10^{-8}$ and $\omega_h = 10^{-2}$. The numerical results are shown in Fig. S17. Similar to the Markovian noise, $\bar{E}_A$ also shows a linear increase in a short time regime, see Fig. S17a. However, $1/f$-noise-induced information leakage depends on the initial state, despite the same energy, see Fig. S17b. As shown in Fig. S16c the slopes of $\bar{E}_A(t)$ are proportional to $\sqrt{A_j}$ (which is similar to single-qubit dephasing time $T_2^*$). Therefore, $\bar{E}_A$ **serves as an effective measure of many-body information leakage for both Markovian and $1/f$ noise.**

## Appendix G: Particle propagation

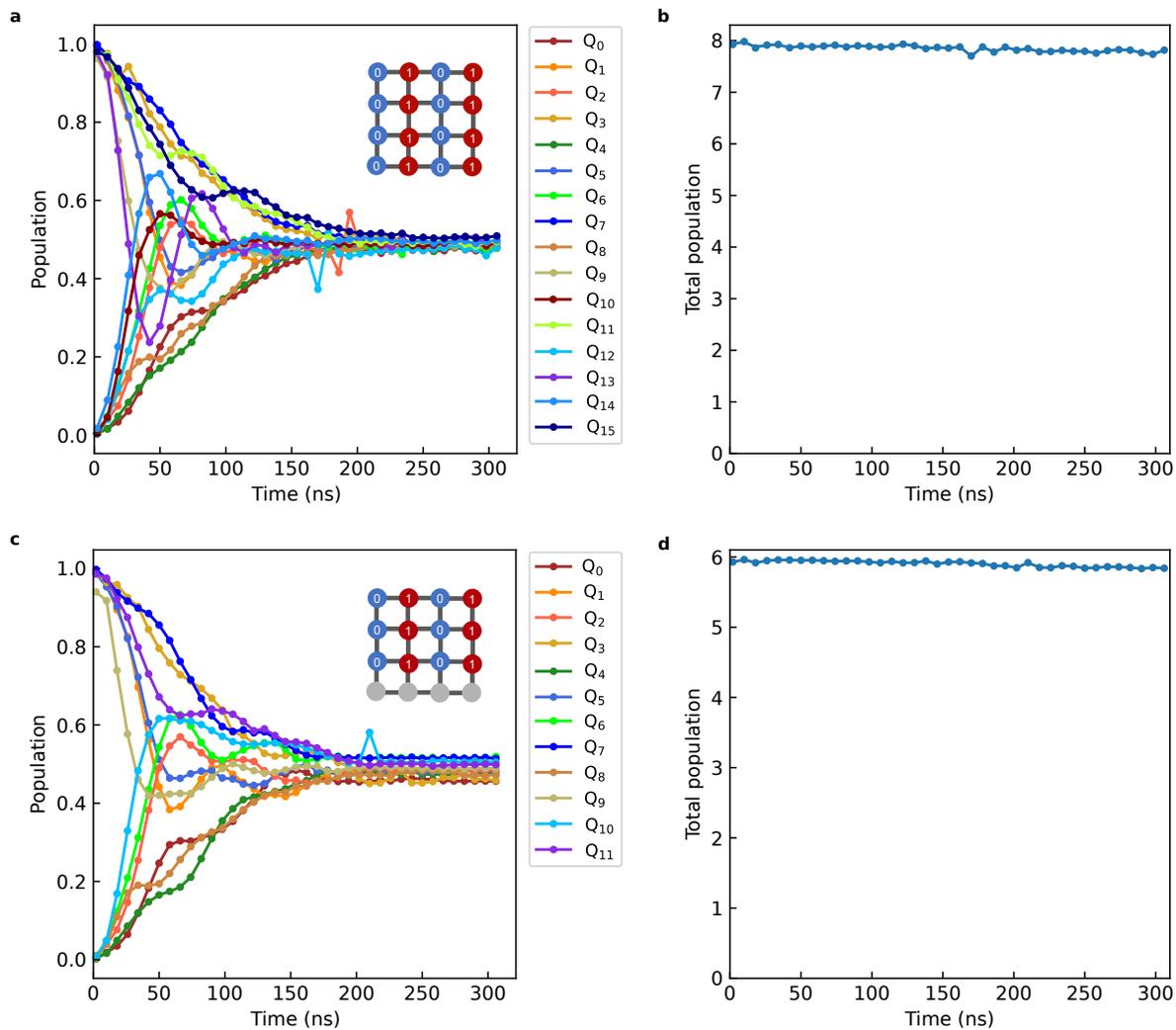

FIG. S18. **Particle propagation. a**, Evolution of the Particles in the 16-qubit system with a half–filled initial state. **b**, Total-population evolution in **a**. **c**, Evolution of the particles in the 12-qubit system with a half-filled initial state. **d**, Total-populatoin evolution in **c**.




\end{document}